\documentclass{llncs}

\usepackage{graphicx}

\input{epsf}

\begin{document}

\pagestyle{empty}

\newcommand {\ignore} [1] {}

    \ignore {
\newtheorem{theorem}{Theorem}[section]
\newtheorem{lemma}[theorem]{Lemma}
\newtheorem{corollary}[theorem]{Corollary}
\newtheorem{definition}[theorem]{Definition}
\newtheorem{proposition}[theorem]{Proposition}
\newtheorem{fact}[theorem]{Fact}
\newenvironment{proof}{\noindent{\bf Proof:\/}}{$\Box$\vskip 0.1in}
\newenvironment{proofsp}{\noindent{\bf Proof}}{$\Box$\vskip 0.1in}
\newenvironment{prooffree}{\noindent}{$\Box$\vskip 0.1in}
}

\def \bfm {\bf \boldmath}

\def \AA{{\cal A}}
\def \BB{{\cal B}}
\def \CC{{\cal C}}
\def \DD{{\cal D}}
\def \EE{{\cal E}}
\def \FF{{\cal F}}
\def \GG{{\cal G}}\def \HH{{\cal H}}
\def \II{{\cal I}}
\def \JJ{{\cal J}}
\def \KK{{\cal K}}
\def \LL{{\cal L}}
\def \MM{{\cal M}}
\def \NN{{\cal N}}
\def \OO{{\cal O}}
\def \PP{{\cal P}}
\def \QQ{{\cal Q}}
\def \RR{{\cal R}}
\def \SS{{\cal S}}
\def \TT{{\cal T}}
\def \UU{{\cal U}}
\def \VV{{\cal V}}
\def \WW{{\cal W}}
\def \XX{{\cal X}}
\def \YY{{\cal Y}}
\def \ZZ{{\cal Z}}

\def \ee{{\varepsilon}}

\mainmatter

\title{
%Towards
 Restructuring in Combinatorial Optimization
%design of Spanning Trees
}

\author{Mark Sh. Levin}

\institute{Inst. for Information Transmission Problems,
 Russian Acad. of Sci., \\
 19 Bolshoj Karetny lane, Moscow 127994, Russia\\
 \email{mslevin@acm.org}}
%

%\normalsize
\maketitle

\begin {abstract}

% Reengineering (redesign) of existing modular systems is a
% significant systems engineering problem.
 The paper addresses
 a new class of combinatorial problems
 which consist in restructuring of solutions
 (as structures)
 in combinatorial optimization.
 Two main features of the restructuring process are examined:
 (i) a cost of the restructuring,
 (ii) a closeness to a goal solution.
 This problem corresponds to redesign (improvement, upgrade) of modular
 systems or solutions.
 The restructuring approach is described and illustrated for the following combinatorial
 optimization problems:
 knapsack problem,
 multiple choice problem,
 assignment problem,
 spanning tree problems.
%
% Numerical
 Examples
 illustrate the restructuring processes.

~~


\end{abstract}

\newcounter{cms}
\setlength{\unitlength}{1mm}

%\tableofcontents

\section{Introduction}

 % Reengineering (redesign) of existing modular systems is a
% significant systems engineering problem.
 The paper addresses
 a new class of combinatorial problems
 which are targeted to
 restructuring of solutions (e.g., a set of elements, a structure)
 in combinatorial optimization.
 Two main features of the restructuring process are examined:
 (i) a cost of the initial problem solution restructuring,
 (ii) a closeness the obtained restructured solution to a goal solution.
 Fig. 1 depicts the restructuring process.

\begin{center}
\begin{picture}(80,52)

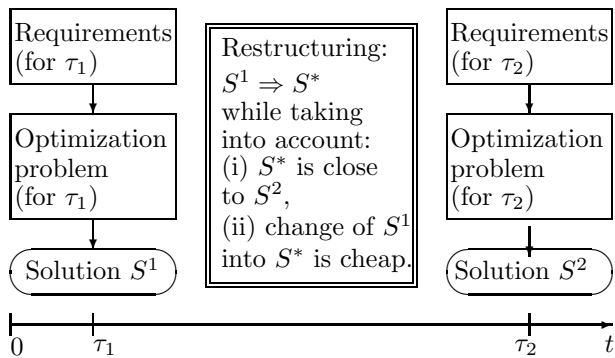
\put(06,00){\makebox(0,0)[bl]{Fig. 1. Illustration for
restructuring process}}

\put(00,09){\vector(1,0){80}}

\put(00,7.5){\line(0,1){3}} \put(11,7.5){\line(0,1){3}}
\put(69,7.5){\line(0,1){3}}

\put(00,05){\makebox(0,0)[bl]{\(0\)}}
\put(11,05){\makebox(0,0)[bl]{\(\tau_{1}\)}}
\put(67,05){\makebox(0,0)[bl]{\(\tau_{2}\)}}

\put(79,05.3){\makebox(0,0)[bl]{\(t\)}}

%=================================== S1

\put(00,41){\line(1,0){22}} \put(00,51){\line(1,0){22}}
\put(00,41){\line(0,1){10}} \put(22,41){\line(0,1){10}}

\put(0.5,46){\makebox(0,0)[bl]{Requirements}}
\put(0.5,42){\makebox(0,0)[bl]{(for \(\tau_{1}\))}}

%--

\put(11,41){\vector(0,-1){4}}

\put(00,23){\line(1,0){22}} \put(00,37){\line(1,0){22}}
\put(00,23){\line(0,1){14}} \put(22,23){\line(0,1){14}}

\put(0.5,32){\makebox(0,0)[bl]{Optimization}}
\put(0.5,28){\makebox(0,0)[bl]{problem}}
\put(0.5,24){\makebox(0,0)[bl]{(for \(\tau_{1})\)}}

%--

\put(11,23){\vector(0,-1){4}}

\put(11,16){\oval(22,06)}

\put(02,15){\makebox(0,0)[bl]{Solution \(S^{1}\)}}

%%%%%%%%%%%%%%%%%%%%%%%%%%%%%%%%%%%%%%%%%%%%%%%%%%%%%%%%%%

\put(26,14){\line(1,0){28}} \put(26,49){\line(1,0){28}}
\put(26,14){\line(0,1){35}} \put(54,14){\line(0,1){35}}

\put(26.5,14.5){\line(1,0){27}} \put(26.5,48.5){\line(1,0){27}}
\put(26.5,14.5){\line(0,1){34}} \put(53.5,14.5){\line(0,1){34}}

\put(28,44){\makebox(0,0)[bl]{Restructuring:}}
\put(28,40){\makebox(0,0)[bl]{\(S^{1} \Rightarrow S^{*}  \)}}
\put(28,35.5){\makebox(0,0)[bl]{while taking}}
\put(28,32.5){\makebox(0,0)[bl]{into account:}}
\put(28,28){\makebox(0,0)[bl]{(i) \(S^{*}\) is close}}
\put(28,24.4){\makebox(0,0)[bl]{to \(S^{2}\),}}
\put(28,20){\makebox(0,0)[bl]{(ii) change of \(S^{1}\)}}
\put(28,16){\makebox(0,0)[bl]{into \(S^{*}\) is cheap.}}
%\put(28,22){\makebox(0,0)[bl]{is cheap.}}

%%%%%%%%%%%%%%%%%%%%%%%%%%%%%%%%%%%%%%%%%%%%%%%%%%%%%%%%%%

%================================== S2

%--

\put(58,41){\line(1,0){22}} \put(58,51){\line(1,0){22}}
\put(58,41){\line(0,1){10}} \put(80,41){\line(0,1){10}}

\put(58.5,46){\makebox(0,0)[bl]{Requirements}}
\put(58.5,42){\makebox(0,0)[bl]{(for \(\tau_{2}\))}}

%--

\put(69,41){\vector(0,-1){4}}

\put(58,23){\line(1,0){22}} \put(58,37){\line(1,0){22}}
\put(58,23){\line(0,1){14}} \put(80,23){\line(0,1){14}}

\put(58.5,32){\makebox(0,0)[bl]{Optimization}}
\put(58.5,28){\makebox(0,0)[bl]{problem }}
\put(58.5,24){\makebox(0,0)[bl]{(for \(\tau_{2})\)}}

%--

\put(69,22){\vector(0,-1){4}}

\put(69,16){\oval(22,06)}

\put(59,15){\makebox(0,0)[bl]{Solution \(S^{2}\)}}

%===================================

\end{picture}
\end{center}

  % This problem corresponds to redesign (improvement, upgrade) of modular
% systems or solutions.
%
 This kind of problems corresponds to
 redesign/reconfiguration (improvement, upgrade)
 of modular systems and
 the situations can be faced
 in complex software, algorithm systems,
 communication networks, computer networks,
 information systems, manufacturing systems,
 constructions,
%  civil engineering,
 etc.
%
% \cite{ahlund03},
%  (\cite{bal08},
     (\cite{arain08},
      \cite{bi08},
       \cite{bon02},
%     \cite{lev98},
     \cite{lev06},
      \cite{lev09},
      \cite{lev10a},
%
% \cite{lev07},
%     \cite{levdan05},
%     \cite{levsaf06},
%   \cite{nolt99},
%    \cite{rouse05},
     \cite{vrba10}).
  Here an optimization problem is solved for two time moments:
 \(\tau_{1}\) and  \(\tau_{2}\) to obtain corresponding solutions
 \(S^{1}\) and \(S^{2}\).
 The examined restructuring problem consists in
 a ``cheap'' transformation (change) of solution \(S^{1}\) to a solution \(S^{*}\) that
 is very close to \(S^{2}\).
 This restructuring approach is described and illustrated for the following combinatorial
 optimization problems
 (e.g., \cite{gar79},  \cite{lev09}):
 knapsack problem,
 multiple choice problem,
 assignment problem,
 spanning tree problems.
 Numerical examples
 illustrate the restructuring processes.
%
%
% Some Applied examples are briefly described.

%%%%%%%%%%%%%%%%%%%%%%%%%%%%%%%%%%%%%%%%%%%%%%%%%%%%%%%%%%%%%%%

%>>>>>>>>>>>>>>>>>>>>>>>>>>>>>>>>>>>>>>>>>>>>>>>>>>>>>>>>>>>>>>>>>>>>>>>>>>>>>>>>>>
%===================================================

\section{General Restructuring Problems}

%% A prospective problem kind of problem consists in
%% restructuring (reconfiguration, resolving) (e.g., \cite{lev09}).
%
 The restructuring problem may be used for many combinatorial
 optimization problems as changing a solution (e.g., subset, structure),
 for example:
 (i) ranking (sorting) problem,
 (ii) knapsack problem,
 (iii) multiple choice problem,
 (iv) clustering problem,
 (v) assignment/allocation problems,
 (vi) bin-packing problem,
 (vii) graph coloring problem,
 (viii) vertex covering problems,
 (ix) clique problem,
 (x) spanning tree problem, and
%  (Fig. 6),
%
 (xi) Steiner problem.
%  (Fig. 7).
%
 Here it is necessary to take into account
 a cost of solution changes
 (e.g., removal of a Steiner node).
 Fig. 2 illustrates the restructuring problem.

\begin{center}
\begin{picture}(73,49)
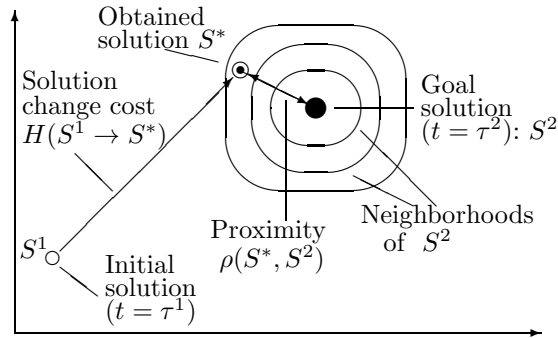
\put(01,00){\makebox(0,0)[bl]{Fig. 2. Illustration for
restructuring problem}}

\put(00,05){\vector(0,1){43}} \put(00,05){\vector(1,0){70}}

%----------------S1

\put(0.6,15){\makebox(0,0)[bl]{\(S^{1}\)}}
\put(5,15){\circle{1.7}}

\put(12,13){\makebox(0,0)[bl]{Initial}}
\put(12,10){\makebox(0,0)[bl]{solution}}
\put(12,06){\makebox(0,0)[bl]{(\(t=\tau^{1}\))}}

\put(11,10){\line(-4,3){5}}

\put(6,16){\vector(1,1){23}}

%----------------S2

\put(40,35){\circle*{2.7}}

\put(54,37){\makebox(0,0)[bl]{Goal}}
\put(54,34){\makebox(0,0)[bl]{solution}}
\put(54,30){\makebox(0,0)[bl]{(\(t=\tau^{2}\)): \(S^{2}\)}}
%\put(54,30){\makebox(0,0)[bl]{\(S^{2}\)}}

\put(53,35){\line(-1,0){10.5}}

\put(40,35){\oval(12,10)} \put(40,35){\oval(17,17)}
\put(40,35){\oval(24,22)}

%++++++++++++++++++++++++++++++Proximity

\put(40,35){\vector(-2,1){9}} \put(30,40){\vector(2,-1){9}}

%\put(40,35){\line(-2,1){10}}

%--

\put(26,17){\makebox(0,0)[bl]{Proximity}}
\put(26,13){\makebox(0,0)[bl]{~\(\rho (S^{*},S^{2})\)}}

\put(36,20){\line(0,1){16}}

%++++++++++++++++++++++++++++++
%------------------------------Neighborhoods of S2

\put(46,19){\makebox(0,0)[bl]{Neighborhoods }}
\put(49,16){\makebox(0,0)[bl]{of ~\(S^{2}\)}}

\put(56,22){\line(-1,1){10}}

\put(53,22){\line(-2,1){8}}

%++++++++++++++++++++++++++++++

%----------------S(xi) - obtained solution

\put(30,40){\circle{2}} \put(30,40){\circle*{1}}

\put(11,46){\makebox(0,0)[bl]{Obtained}}
\put(11,43){\makebox(0,0)[bl]{solution \(S^{*}\)}}

\put(20,42.8){\line(4,-1){7}}

%--------------------------------------
\put(1,37.5){\makebox(0,0)[bl]{Solution }}
\put(1,33.5){\makebox(0,0)[bl]{change cost }}
\put(1,29.5){\makebox(0,0)[bl]{\(H(S^{1} \rightarrow S^{*})\)}}

\put(8,29){\line(1,-1){5}}

%--------------------------------------

\end{picture}
\end{center}

 Let \(P\) be a combinatorial optimization problem with a solution as
 structure
 \(S\)
 (i.e., subset, graph),
 \(\Omega\) be initial data (elements, element parameters, etc.),
 \(f(P)\) be objective function(s).
 Thus \(S(\Omega)\) be a solution for initial data \(\Omega\),
 \(f(S(\Omega))\) be the corresponding objective function.
 Let \(\Omega^{1}\) be initial data at an initial stage,
  \(f(S(\Omega^{1}))\) be the corresponding objective function.
 \(\Omega^{2}\) be initial data at next stage,
  \(f(S(\Omega^{2}))\) be the corresponding objective function.
 As a result,
 the following solutions can be considered:
 ~(a) \( S^{1}=S(\Omega^{1})\) with \(f(S(\Omega^{1}))\) and
 ~(b) \( S^{2}=S(\Omega^{2})\) with \(f(S(\Omega^{2}))\).
 In addition it is reasonable to examine a cost of changing
 a solution into another one:~
 \( H(S^{\alpha} \rightarrow  S^{\beta})\).
 Let \(\rho ( S^{\alpha}, S^{\beta} )\) be a proximity between solutions
  \( S^{\alpha}\) and \( S^{\beta}\),
  for example,
 \(\rho ( S^{\alpha}, S^{\beta} ) = | f(S^{\alpha}) -  f(S^{\beta}) |\).
 Note function \(f(S)\) is often a vector function.
 Finally, the restructuring problem can be examine as follows (a basic version):

~~

 Find a solution \( S^{*}\) while taking into account the
 following:

  (i) \( H(S^{1} \rightarrow  S^{*}) \rightarrow \min \),
%  ~ and
%
  ~(ii) \(\rho ( S^{*}, S^{2} )  \rightarrow \min  \) ~(or constraint).

~~

 Thus the basic optimization model can be examined as the following:

% {\it basic restructuring problem 1}:~~
%
   \[\min \rho ( S^{*}, S^{2} ) ~~~s.t.
 ~ H(S^{1} \rightarrow  S^{*})  \leq \widehat{h}, \]
 where \(\widehat{h}\) is a constraint for cost of the solution
 change.
%
%%% {\it basic restructuring problem 2}:~~
%
%%%   \(\min H(S^{1} \rightarrow  S^{*})  ~~~ s.t.
%
%%%  ~ \rho ( S^{*}, S^{2} )  \leq \rho^{0} \).
%
%%%%%%%%%%%%%%%%%%%%%%%%%%%%%%%%%%%%%%%%%%%%%%%%%%%%%%%%%%%%%%%%%%%%%%%%%%%%%%%%%
%%%%%%%%%%%%%%%%%%%%%%%%%%%%%%%%%%%%%%%%%%%%%%%%%%%%%%%%%%%%%%%%%%%%%%%%%%%%%%%%%
%%%%%%%%%%%%%%%%%%%%%%%%%%%%%%%%%%%%%%%%%%%%%%%%%%%%%%%%%%%%%%%%%%%%%%%%%%%%%%%%%
%
 Fig. 3 illustrates restructuring of a multicriteria  problem.
 Note proximity function  ~\(\rho (S^{*},S^{2}) \)~
 (or  ~\(\rho (S^{*j}, \{S^{21},S^{22},S^{23}\} \))
  can be considered as a
 vector function as well
%  (e.g., \cite{lev98})
 (analogically for the solution change cost).
 This situation will lead to a multicriteria restructuring problem.

\begin{center}
\begin{picture}(80,55)
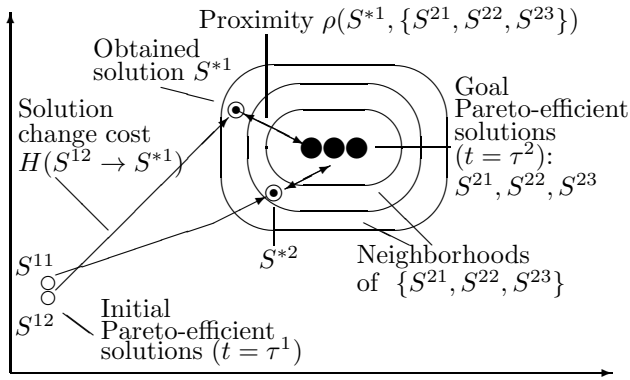
\put(05,00){\makebox(0,0)[bl]{Fig. 3. Restructuring of
multicriteria problem}}

\put(00,05){\vector(0,1){48}} \put(00,05){\vector(1,0){80}}

%----------------S1

\put(0.6,18){\makebox(0,0)[bl]{\(S^{11}\)}}
\put(0.6,10){\makebox(0,0)[bl]{\(S^{12}\)}}

\put(5,17){\circle{1.7}} \put(5,15){\circle{1.7}}

\put(12,12.5){\makebox(0,0)[bl]{Initial}}
\put(12,09.5){\makebox(0,0)[bl]{Pareto-efficient}}
\put(12,06){\makebox(0,0)[bl]{solutions (\(t=\tau^{1}\))}}
%\(S^{11},S^{12}\)}}

\put(11,10){\line(-4,3){4}}

\put(6,16){\vector(1,1){23}}

\put(6,18){\line(3,1){21}} \put(27,25){\vector(2,1){07}}

%%%%%%Obtained solution 2: S(*2)
\put(35,29){\circle{2}} \put(35,29){\circle*{1}}

\put(33,19){\makebox(0,0)[bl]{\(S^{*2}\)}}
\put(35,22){\line(0,1){5}}

%-- proximity S*2 - S2

\put(36.5,29.5){\vector(2,1){6}}
\put(42.5,32.5){\vector(-2,-1){6}}

%----------------S2

\put(40,35){\circle*{2.7}} \put(43,35){\circle*{2.7}}
\put(46,35){\circle*{2.7}}

\put(59,42){\makebox(0,0)[bl]{Goal}}
\put(59,39){\makebox(0,0)[bl]{Pareto-efficient}}
\put(59,36){\makebox(0,0)[bl]{solutions}}
\put(59,32){\makebox(0,0)[bl]{(\(t=\tau^{2}\)):}}
\put(59,28){\makebox(0,0)[bl]{\(S^{21},S^{22},S^{23}\)}}

\put(58.5,35){\line(-1,0){10}}

\put(43,35){\oval(18,10)} \put(43,35){\oval(23,17)}
\put(43,35){\oval(30,22)}

%++++++++++++++++++++++++++++++Proximity

\put(40,35){\vector(-2,1){9}} \put(30,40){\vector(2,-1){9}}

%\put(40,35){\line(-2,1){10}}

%--

%\put(32,54){\makebox(0,0)[bl]{Proximity}}
%\put(26,50){\makebox(0,0)[bl]{~\(\rho
%(S^{*1},\{S^{21},S^{22},S^{23}\})\)}}

%\put(32,54){\makebox(0,0)[bl]{Proximity}}
\put(25,50){\makebox(0,0)[bl]{Proximity \(\rho
(S^{*1},\{S^{21},S^{22},S^{23}\})\)}}

\put(34,50){\line(0,-1){11}}

%++++++++++++++++++++++++++++++
%------------------------------Neighborhoods of S2

\put(46,19){\makebox(0,0)[bl]{Neighborhoods }}
\put(46,15){\makebox(0,0)[bl]{of ~\(\{S^{21},S^{22},S^{23}\}\)}}

\put(56,22){\line(-1,1){7}}

\put(53,22){\line(-2,1){6}}

%++++++++++++++++++++++++++++++

%----------------S(*) - obtained solution

\put(30,40){\circle{2}} \put(30,40){\circle*{1}}

\put(11,47){\makebox(0,0)[bl]{Obtained}}
\put(11,44){\makebox(0,0)[bl]{solution \(S^{*1}\)}}

\put(20,43){\line(4,-1){7}}

%--------------------------------------
\put(1,39){\makebox(0,0)[bl]{Solution }}
\put(1,35){\makebox(0,0)[bl]{change cost }}
\put(1,31){\makebox(0,0)[bl]{\(H(S^{12} \rightarrow S^{*1})\)}}

\put(6,31){\line(1,-1){7}}

%--------------------------------------

%%%%%%%%%%%%%%%%%%%%%%%%%%%%%%%%%%%%%%%%%%%%%%%%%%%%%%%%%%%%%%

\end{picture}
\end{center}

%=============================================================

\section{Restructuring in Some Combinatorial
%Optimization
 Problems}

%%% In this section the restructuring  process
%%% (or solution reconfiguration) is applied
%%% to some combinatorial optimization problems.
%
% (a) knapsack problem,
%
% (b) multiple choice problem, and
%
% (c) assignment problem, and
%
% (d) spanning tree problems.
%
% Generally, problem descriptions and formulations are described.

%\subsection{Knapsack Problem}

 Let \(A=\{1,...,i,...,n\}\) be an initial set of elements.
  Knapsack problem is considered
 for two time moments \(\tau_{1}\) and \(\tau_{2}\)
 (for \(\tau_{2}\) parameters \(\{c^{2}_{i}\}\), \(\{a^{2}_{i}\}\), and \(b^{2}\) are used)
 (Fig. 4):

% (\cite{gar79}, \cite{keller04}
% , \cite{mar90}

\begin{center}
\begin{picture}(80,55)

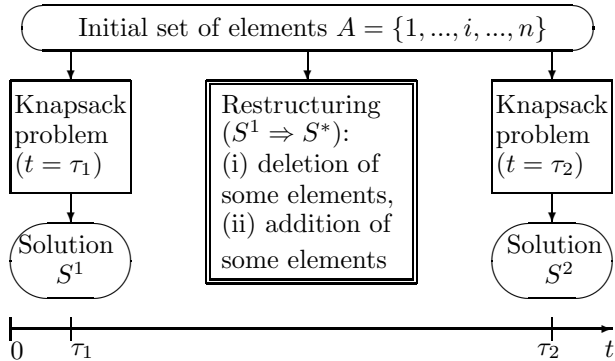
\put(07,00){\makebox(0,0)[bl]{Fig. 4. Restructuring in knapsack
problem}}

\put(00,10){\vector(1,0){80}}

\put(00,8.5){\line(0,1){3}} \put(8,8.5){\line(0,1){3}}
\put(72,8.5){\line(0,1){3}}

\put(00,06){\makebox(0,0)[bl]{\(0\)}}
\put(8,06){\makebox(0,0)[bl]{\(\tau_{1}\)}}
\put(70,06){\makebox(0,0)[bl]{\(\tau_{2}\)}}

%\put(79,06){\makebox(0,0)[bl]{\(t\)}}

\put(79,06.3){\makebox(0,0)[bl]{\(t\)}}

%=================================== S1

%\put(00,47){\line(1,0){22}} \put(00,58){\line(1,0){22}}
%\put(00,47){\line(0,1){11}} \put(22,47){\line(0,1){11}}

%\put(0.5,53){\makebox(0,0)[bl]{Requirements}}
%\put(0.5,49){\makebox(0,0)[bl]{(for \(\tau_{1}\))}}

%--

%\put(11,47){\vector(0,-1){4}}

\put(00,28){\line(1,0){16}} \put(00,43){\line(1,0){16}}
\put(00,28){\line(0,1){15}} \put(16,28){\line(0,1){15}}

\put(0.5,38){\makebox(0,0)[bl]{Knapsack}}
\put(0.5,34){\makebox(0,0)[bl]{problem}}
\put(0.5,30){\makebox(0,0)[bl]{(\(t=\tau_{1})\)}}

%--

\put(8,28){\vector(0,-1){4}}

\put(8,19){\oval(16,10)}

\put(01,20){\makebox(0,0)[bl]{Solution}}

\put(06,16){\makebox(0,0)[bl]{\(S^{1}\)}}

%%%%%%%%%%%%%%%%%%%%%%%%%%%%%%%%%%%%%%%%%%%%%%%%%%%%%%%%%%

\put(26,16){\line(1,0){28}} \put(26,43){\line(1,0){28}}
\put(26,16){\line(0,1){27}} \put(54,16){\line(0,1){27}}

\put(26.5,16.5){\line(1,0){27}} \put(26.5,42.5){\line(1,0){27}}
\put(26.5,16.5){\line(0,1){26}} \put(53.5,16.5){\line(0,1){26}}

%\put(28,50){\makebox(0,0)[bl]{Resolving:}}
%\put(28,46){\makebox(0,0)[bl]{\(S^{1} \Rightarrow S^{*}  \)}}
%\put(28,42){\makebox(0,0)[bl]{Requirements:}}
\put(28,38){\makebox(0,0)[bl]{Restructuring}}
\put(28,34){\makebox(0,0)[bl]{(\(S^{1} \Rightarrow S^{*}\)):}}
\put(28,30){\makebox(0,0)[bl]{(i) deletion of }}
\put(28,26){\makebox(0,0)[bl]{some elements,}}
\put(28,22){\makebox(0,0)[bl]{(ii) addition of }}
\put(28,18){\makebox(0,0)[bl]{some elements}}

%------------------------

\put(39.5,50){\oval(76,06)}

\put(09,48){\makebox(0,0)[bl]{Initial set of elements \(A =
\{1,...,i,...,n\}\)}}

%\put(31,46){\makebox(0,0)[bl]{\(A = \{1,...,i,...,n\} \)}}

\put(8,47){\vector(0,-1){4}} \put(39.5,47){\vector(0,-1){4}}
\put(72,47){\vector(0,-1){4}}

%%%%%%%%%%%%%%%%%%%%%%%%%%%%%%%%%%%%%%%%%%%%%%%%%%%%%%%%%%

%================================== S2

%\put(58,47){\line(1,0){22}} \put(58,58){\line(1,0){22}}
%\put(58,47){\line(0,1){11}} \put(80,47){\line(0,1){11}}

%\put(58.5,53){\makebox(0,0)[bl]{Requirements}}
%\put(58.5,49){\makebox(0,0)[bl]{(for \(\tau_{2}\))}}

%--

%\put(69,47){\vector(0,-1){4}}

\put(64,28){\line(1,0){16}} \put(64,43){\line(1,0){16}}
\put(64,28){\line(0,1){15}} \put(80,28){\line(0,1){15}}

\put(64.5,38){\makebox(0,0)[bl]{Knapsack }}
\put(64.5,34){\makebox(0,0)[bl]{problem }}
\put(64.5,30){\makebox(0,0)[bl]{(\(t=\tau_{2})\)}}

%--

\put(72,28){\vector(0,-1){4}}

\put(72,19){\oval(16,10)}

\put(66,20){\makebox(0,0)[bl]{Solution}}

\put(71,16){\makebox(0,0)[bl]{\(S^{2}\)}}

%===================================

\end{picture}
\end{center}

%%%%%%%%%%%%%%%%%%%%%%%%%%%%%%%%%%%%%%%%%%%%%%%%%%%%%%%%%%%%%%%

%%%%%%%%%%%%%%%%%%%%%%%%%%%%%%%%%%%%%%%%%%%%%%%%%%%%%%%%%%
%
 \[\max\sum_{i=1}^{n} c^{1}_{i} x_{i}
 ~~~ s.t.~ \sum_{i=1}^{n} a^{1}_{i} x_{i} ~\leq~ b^{1},
 ~~~ x_{i} \in \{0,1\}.  \]
%
%and
%%%%%%%%%%%%%%%%%%%%%%%%%%%%%%%%%%%%%%%%%%%%%%%%%%%%%%%%%%
%
%%% \[\max\sum_{i=1}^{n} c^{2}_{i} x_{i}
%
%%% ~~~ s.t.~ \sum_{i=1}^{n} a^{2}_{i} x_{i} ~\leq~ b^{2},
%
%%% ~~~ x_{i} \in \{0,1\}.  \]
%
 The corresponding solutions are:
 \(S^{1} \subseteq A \)
% for
 (\(t=\tau_{1}\))
 and
 \(S^{2} \subseteq A \)
% for
 (\(t=\tau_{2}\))
 (\(S^{1} \neq S^{2}\)).

 {\bf Illustrative numerical example:}~
 \(A = \{1,2,3,4,5,6,7\}\),
 \(S^{1} = \{1,3,4,5\}\),
 \(S^{2} = \{2,3,5,7\}\),
  \(S^{*} = \{2,3,4,6\}\).
  The change (restructuring) process (i.e., \(S^{1} \Rightarrow S^{*}\)) is based
  on the following (Fig. 5):
 (a) deleted elements:
%   (\( S^{*} \backslash S^{1} \)   )    :
 \( S^{1*-} =  S^{1} \backslash S^{*} =  \{1,5\}\),
 (b) added elements:
  \(S^{1*+} = S^{*} \backslash S^{1} = \{2,6\}\).

\begin{center}
\begin{picture}(80,37)

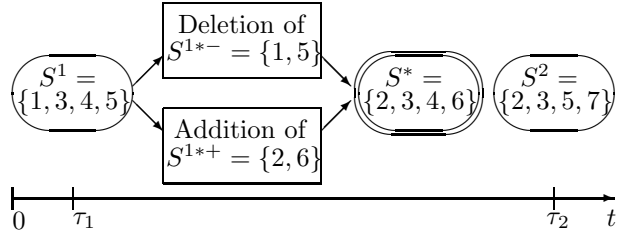
\put(014,00){\makebox(0,0)[bl]{Fig. 5. Example for restructuring}}

\put(00,10){\vector(1,0){80}}

\put(00,8.5){\line(0,1){3}} \put(8,8.5){\line(0,1){3}}
\put(72,8.5){\line(0,1){3}}

\put(00,06){\makebox(0,0)[bl]{\(0\)}}
\put(8,06){\makebox(0,0)[bl]{\(\tau_{1}\)}}
\put(71,06){\makebox(0,0)[bl]{\(\tau_{2}\)}}

%\put(79,06){\makebox(0,0)[bl]{\(t\)}}

\put(79,06.3){\makebox(0,0)[bl]{\(t\)}}

%================================== S1

\put(16,25){\vector(1,1){4}} \put(16,23){\vector(1,-1){4}}

\put(8,24){\oval(16,10)}

\put(3.5,25){\makebox(0,0)[bl]{\(S^{1} =\)}}
\put(0.5,21){\makebox(0,0)[bl]{\(\{1,3,4,5\}\)}}

%===================================

%%%%%%%%%%%%%%%%%%%%%%%%%%%%%%%%%%%%%%%%%%%%%%%%%%%%%%%%%%

\put(20,26){\line(1,0){21}} \put(20,36){\line(1,0){21}}
\put(20,26){\line(0,1){10}} \put(41,26){\line(0,1){10}}

%\put(20.5,26.5){\line(1,0){20}} \put(20.5,35.5){\line(1,0){20}}
%\put(20.5,26.5){\line(0,1){09}} \put(40.5,26.5){\line(0,1){09}}

\put(22,32){\makebox(0,0)[bl]{Deletion of }}
\put(20.4,27.5){\makebox(0,0)[bl]{\(S^{1*-}=\{1,5\}\)}}

%--

\put(20,12){\line(1,0){21}} \put(20,22){\line(1,0){21}}
\put(20,12){\line(0,1){10}} \put(41,12){\line(0,1){10}}

%\put(20.5,12.5){\line(1,0){20}} \put(20.5,21.5){\line(1,0){20}}
%\put(20.5,12.5){\line(0,1){09}} \put(40.5,12.5){\line(0,1){09}}

\put(21.4,18){\makebox(0,0)[bl]{Addition of }}
\put(20.4,13.5){\makebox(0,0)[bl]{\(S^{1*+}=\{2,6\}\)}}

%------------------------

%%\put(39.5,50){\oval(76,06)}

%%\put(09,48){\makebox(0,0)[bl]{Initial set of elements \(A =
%%\{1,...,i,...,n\}\)}}

%\put(31,46){\makebox(0,0)[bl]{\(A = \{1,...,i,...,n\} \)}}

%%\put(8,47){\vector(0,-1){4}} \put(39.5,47){\vector(0,-1){4}}
%%\put(72,47){\vector(0,-1){4}}

%%%%%%%%%%%%%%%%%%%%%%%%%%%%%%%%%%%%%%%%%%%%%%%%%%%%%%%%%%

\put(41,29){\vector(1,-1){4}} \put(41,19){\vector(1,1){4}}

%================================== S*

%\put(71,28){\vector(0,-1){4}}

\put(54,24){\oval(16,10)} \put(54,24){\oval(17,11)}

\put(49.5,25){\makebox(0,0)[bl]{\(S^{*} =\)}}
\put(46.5,21){\makebox(0,0)[bl]{\(\{2,3,4,6\}\)}}

%================================== S2

%\put(71,28){\vector(0,-1){4}}

\put(72,24){\oval(16,10)}

\put(67.5,25){\makebox(0,0)[bl]{\(S^{2} =\)}}
\put(64.5,21){\makebox(0,0)[bl]{\(\{2,3,5,7\}\)}}

%===================================

\end{picture}
\end{center}

 Note the following exists
 at the start stage of the solving process:
 ~\( S^{1*-} = S^{1} \) and
% ~\( S^{1+} = S^{2} \backslash S^{1} \).
%
 \( S^{1*+} =
% (S^{2} \backslash  S^{1})  \bigcup
 A \backslash  S^{1}   \).
%
%
% Thus
 The restructuring problem can be considered as the following:
  \[\min \rho ( S^{*} , S^{2})
  ~~~s.t. ~~
   H(S^{1} \Rightarrow S^{*}) = ( \sum_{i \in  S^{1*-} } h^{-}_{i} +  \sum_{i \in S^{1*+}} h^{+}_{i} ~) \leq \widehat{h},
   ~\sum_{i \in S^{*}}  a^{2}_{i} \leq b^{2}, \]
 where \(\widehat{h}\) is a constraint for the change cost,
 \(h^{-}(i)\) is a cost of deletion of element \(i \in A\), and
  \(h^{+}(i)\) is a cost of addition of element \(i \in A\).
  On the other hand, an equivalent problem can be examined:
 \[\max \sum_{i \in S^{*}} x_{i} c_{i}^{2}
  ~~~s.t. ~~~
  H ( S^{1} \Rightarrow S^{*} ) =
  ( \sum_{i \in S^{1*-}} h^{-}_{i} + \sum_{i \in S^{1*+}} h^{+}_{i} ~) \leq \widehat{h},
   ~\sum_{i \in S^{*} }  a^{2}_{i} \leq b^{2},\]
 because
  ~\( \max \sum_{i \in S^{*}} x_{i} c_{i}^{2}
   \leq
     \max \sum_{i \in {S}^{2}} x_{i} c_{i}^{2} \)~
 while taking into account constraint:
   ~\( \sum_{i \in {S}^{*}}  a^{2}_{i} \leq b^{2} \).
%
%%%%%%%%%%%%%%%%%%%%%%%%%%%%%%%%%%%%%%%%%%%%%%%%%%%%%%%%%%%%%%%%%%%%%%%%%%%
%
  The obtained problem is a modified knapsack-like problem as well.
  At the same time, it is possible to use a simplified solving
  scheme (by analysis of {\it change elements} for addition/deletion):
 (a) generation  of candidate elements for deletion
 (i.e., selection of \(S^{1-}\) from \(S^{1}\)),
 (b) generation of candidate elements for addition
 (i.e., selection of \(S^{1+}\) from
 \(
% (S^{2} \backslash  S^{1})  \bigcup
 A \backslash  S^{1}  \)).
 The selection processes may be based on multicriteria ranking.
 As a result, a problem with sufficiently
 decreased dimension
 will be obtained.

%%%%%%%%%%%%%%%%%%%%%%%%%%%%%%%%%%%%%%%%%%%%%%%%%%%%%%%%%%%%%%%%%%%%%%%%%%%%%%%%%%%
%\subsection{Multiple Choice Problem}

 Basic multiple choice problem is for \(t=\tau_{1}\)
 (for \(t=\tau_{2}\) parameters \(\{c^{2}_{ij}\}\), \(\{a^{2}_{ij}\}\), and \(b^{2}\)
 are used):
 \[\max\sum_{i=1}^{m} \sum_{j=1}^{q_{i}} c^{1}_{ij} x_{ij}
 ~~~s.t.~\sum_{i=1}^{m} \sum_{j=1}^{q_{i}} a^{1}_{ij} x_{ij} \leq b^{1},
 ~\sum_{j=1}^{q_{i}} x_{ij} \leq 1 ~~ \forall i=\overline{1,m},
 ~~~x_{ij} \in \{0,1\}.\]
 Here initial element set  \(A\) is divided into
 \(m\) subsets (without intersection):
 ~\(A = \bigcup_{i=1}^{m}  A_{i}\), where  ~\(A_{i} =
 \{1,...,j,...,q_{i}\}\) (\(i=\overline{1,m}\)).
 Thus each element is denoted by \((i,j)\).
%
% Thus
 An equivalent problem is:
 \[ \max \sum_{(i,j) \in S^{1}} c_{ij}^{1}~~~~~ s.t.  \sum_{(i,j) \in S^{1}} a_{ij}^{1} \leq b^{1},
  ~~ | S^{1} \& A_{i} | \leq 1 ~ \forall i =  \overline{1,m}. \]
 For \(t=\tau_{2}\) the problem is the same.
%
% \[ \max \sum_{(i,j) \in S^{2}} c_{ij}^{2}~~~~~ s.t.  \sum_{(i,j) \in S^{2}} a_{ij}^{2} \leq b^{2},
%  ~~ | S^{2} \& A_{i} | \leq 1 ~ \forall i =  \overline{1,m}. \]

%
 {\bf Illustrative numerical example:}~
 \(A = \{1,2,3,4,5,6,7,8,9,10,11,12,13\}\),
 \(A_{1} = \{1,3,5,12\}\), \(A_{2} = \{2,7,9\}\),
 \(A_{3} = \{4,8,13\}\), \(A_{4} = \{6,10,11\}\),
 \(S^{1} = \{1,7,8,11\}\),
 \(S^{2} = \{3,7,8,10\}\),
  \(S^{*} = \{1,2,8,6\}\).
  The change (restructuring) process (i.e., \(S^{1} \Rightarrow S^{*}\)) is based
  on the following (Fig. 6):
 (a) deleted elements:
%   (\( S^{*} \backslash S^{1} \)   )    :
 \( S^{1*-} =  S^{1} \backslash S^{*} =  \{7,11\}\),
 (b) added elements:
  \(S^{1*+} = S^{*} \backslash S^{1} = \{2,6\}\).

%%%Note the following exists at the start stage of the solving
%%%process:
%%% ~\( S^{1-} = S^{1} \) and \( S^{1+} = A \backslash  S^{1}   \).
%
%
%
 Thus the restructuring problem can be considered as the following:
  \[\min \rho ( S^{*} , S^{2})\]
  \[s.t. ~~~
   H (S^{1} \Rightarrow S^{*}) = ( \sum_{(i,j) \in  S^{1*-} } h^{-}_{ij} +  \sum_{(i,j) \in S^{1*+}} h^{+}_{ij} ~~) \leq \widehat{h},
   ~~\sum_{(i,j) \in S^{*}}  a^{2}_{ij} \leq b^{2},\]
 \[ | S^{*} \& A_{i} | \leq 1 ~ \forall i =  \overline{1,m}.
   \]
 where \(\widehat{h}\) is a constraint for the change cost,
 \(h^{-}(ij)\) is a cost of deletion of element \((i,j) \in A\), and
  \(h^{+}(ij)\) is a cost of addition of element \((i,j) \in A\).
 An equivalent problem is:
  \[\max  \sum_{(i,j) \in S^{*}}  c^{2}_{ij} \]
  \[s.t. ~~~
   H (S^{1} \Rightarrow S^{*}) = ( \sum_{(i,j) \in  S^{1*-} } h^{-}_{ij} +  \sum_{(i,j) \in S^{1*+}} h^{+}_{ij} ~~) \leq \widehat{h},
   ~~\sum_{(i,j) \in S^{*}}  a^{2}_{ij} \leq b^{2},\]
 \[ | S^{*} \& A_{i} | \leq 1 ~ \forall i =  \overline{1,m}.
   \]
%
% This is a multiple choice problem as well.

%%%%%%%%%%%%%%%%%%%%%%%%%%%%%%%%%%%%%%%%%%%%%%%%%%%%%%%%%%%%%%%%%%%%%%%%%%%%%%%%%%%
%\subsection{Assignment Problem}

 The simplest version of algebraic assignment problem is:
%
%%%%%%%%%%%%%%%%%%%%%%%%%%%%%%%%%%%%%%%%%%%%%%%%%%%%%%%%%%%%%%%%%%%%%%%%%%%%
 \[ \max  \sum_{i=1}^{m} \sum_{j=1}^{n} c^{1}_{i,j} x_{i,j}
%  \[l=\overline{1,m}, ~~k=\overline{1,n}; \]
%
 ~~s.t. \sum_{i=1}^{m} x_{i,j} \leq 1, j=\overline{1,n};
 ~ \sum_{j=1}^{n} x_{i,j} \leq 1, i=\overline{1,m};
 ~ x_{i,j} \in \{0,1\}. \]
%%%%%%%%%%%%%%%%%%%%%%%%%%%%%%%%%%%%%%%%%%%%%%%%%%%%%%%%%%%%%%%%%%%%%%%%%%%%
%
 This problem is polynomially solvable.
 Let us consider \(n=m\).
 Thus a solution can be examined as a permutation of elements
 ~\(A=\{1,...,i,..,n\}\):  \( S = < s[1],...,s[i],..,s[n] > \),
 where \(s[i]\) defines the position of element
 \(i\) in the resultant permutation \(S\).
 Let ~\(c(i,s[i]) \geq 0\)~ (\(i=\overline{1,n}\))~ be a ``profit'' of assignment of element \(i\) into
 position \(s[i]\)
 (i.e., \(\| c(i,s[i]) \|\) is a ``profit'' matrix).

 The combinatorial formulation of assignment problem  is:

~~

 Find permutation \(S\) such that
~
 ~\( \sum_{i=1}^{n} c(i,s[i]) \rightarrow  \max \).

~~

 Now let us consider three solutions (permutations):

 (a) ~\(S^{1} = < s^{1}[1],...,s^{1}[i],..,s^{1}[n] > \)~
% be the solution
%  (i.e., permutation of n elements)
 for \(t=\tau_{1}\),

 (b)
 ~\(S^{2} = < s^{2}[1],...,s^{2}[i],..,s^{2}[n] >\)~
% be the solution
 for \(t=\tau_{2}\),
 and

 (c)
  ~\(S^{*}= < s^{*}[1],...,s^{*}[i],..,s^{*}[n] >\)~  (the restructured
  solution).
%  that is under the search process).
%

%
 {\bf Illustrative numerical example:}~
 \(A = \{1,2,3,4,5,6,7\}\),

 \(S^{1} = \{2,4,5,1,3,7,6\}\),
 ~\(S^{2} = \{4,1,3,7,5,2,6\}\),
 ~\(S^{*} = \{2,4,3,1,5,7,6\}\).

 Here the following changes are made in \(S^{1}\):
 ~\(5 \rightarrow 3 \), \(3 \rightarrow 5 \).
 Clearly, the changes can be based on typical exchange operations:
 {\it 2-exchange}, {\it 3-exchange}, etc.

 Further, let us consider a vector of structural difference
 (by components) for two permutations \(S^{\alpha}\) and \(S^{\beta}\):
  ~\(\{ s^{\alpha}[i] - s^{\beta}[i], i=\overline{1,n} \} \)
  and
  a change cost matrix
  ~\(\| d(i,j) \|\)~ (\(i=\overline{1,n}, j=\overline{1,n}\)).
 Here ~\( d(i,i) =0\) ~ \(\forall i=\overline{1,n}\).
 Evidently, the cost for restructuring
 solution \(S^{1}\) into solution  \(S^{*}\) is:
 ~\( H (S^{1} \rightarrow S^{*}) =  \sum_{i=1}^{n} h(s^{1}[i],s^{*}[i]) \).
 Proximity (by ``profit'') for two permutations \(S^{\alpha}\) and \(S^{\beta}\)
 may be considered as follows:
  \(\rho ( S^{\alpha},S^{\beta}) =
  | \sum_{i=1}^{n} c^{\alpha}(i,s^{\alpha}[i]) - \sum_{i=1}^{n} c^{\beta}(i,s^{\beta}[i])|
  \).
 Finally, the restructuring of assignment is (a simple version):
 \[\min \rho ( S^{*},S^{2} )
 ~~~~s.t.~~ H (S^{1} \rightarrow S^{*}) =  \sum_{i=1}^{n} h(s^{1}[i],s^{*}[i]) \leq \widehat{h}.\]

%%%%%%%%%%%%%%%%%%%%%%%%%%%%%%%%%%%%%%%%%%%%%%%%%%%%%%%%%%%%%%%%%%%%%%%%%
%\subsection{Spanning Trees Problems}

 Restructuring problems for
 minimal spanning tree problem and for Steiner tree problem
 are described as follows (Fig. 6, Fig. 7).
 The following numerical examples are presented:

 {\bf I.} Initial graph (Fig. 6): ~\(G=(A,E)\), where ~\(A=\{1,2,3,4,5,6,7\}\),

 \(E=\{(1,2),(1,4),(1,5),(1,6),(2,3),(2,6),(3,6),(4,5),(4,6),(5,6),(5,7),\)

 \((6,7)\}\).

 {\bf II.} Spanning trees (Fig. 6):

 (i) \(T^{1} = (A,E^{1})\), where
 \(E^{1}=\{(1,2),(1,4),(1,6),(3,5),(5,6),(6,7)\}\),

 (ii) \(T^{2} = (A,E^{2})\), where
 \(E^{2}=\{(1,2),(2,3),(2,6),(4,6),(5,6),(6,7)\}\),

 (iii) \(T^{*} = (A,E^{*})\), where
 \(E^{*}=\{(1,2),(1,4),(2,3),(2,6),(3,5),(6,7)\}\).

  Here the edge changes are (\(T^{1} \rightarrow T^{*}\) as \(E^{1} \rightarrow E^{*}\)):

  \(E^{1*-} = \{(1,6),(5,6)\}\)
  and
  ~\(E^{1*+} = \{(2,3),(2,6)\}\).

\begin{center}
\begin{picture}(95,47)
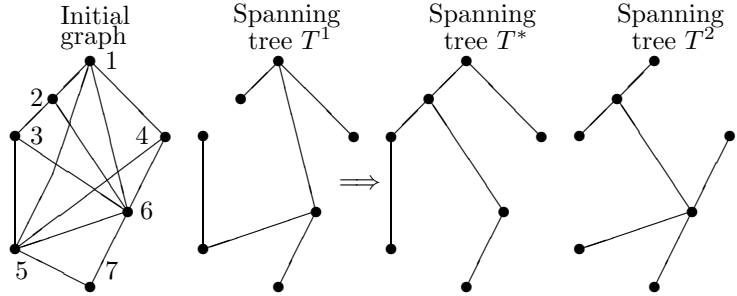
\put(018.5,00){\makebox(0,0)[bl]{Fig. 6. Restructuring of spanning
tree}}

%\put(02,41){\makebox(0,0)[bl]{Initial graph}}

\put(6,42.9){\makebox(0,0)[bl]{Initial}}
\put(6,39.6){\makebox(0,0)[bl]{graph}}

%>>>>>>>>>>>>>>>>>>>>>>>>>>>>>>>>>>>>>>>>>>>>>>> Numbers-labels

\put(12,37){\makebox(0,0)[bl]{\(1\)}}
\put(02,32){\makebox(0,0)[bl]{\(2\)}}
\put(02,27){\makebox(0,0)[bl]{\(3\)}}
\put(16,27){\makebox(0,0)[bl]{\(4\)}}
\put(16.6,17){\makebox(0,0)[bl]{\(6\)}}
\put(00,09){\makebox(0,0)[bl]{\(5\)}}
\put(12,09){\makebox(0,0)[bl]{\(7\)}}

%%%%%%%%%%%%%%%%%%%%%%%%%%%%%%%%%%%%%%%%%%%%%%%%%
%--20

\put(00,28){\circle*{1.3}}

%=======================10-9
\put(00,28){\line(1,1){5}}

\put(00,28){\line(0,-1){15}}

\put(00,28){\line(3,-2){15}}

%%%%%%%%%%%%%%%%%%%%%%%%%%%%%%%%%%%%%%%%%%%%%%%%

%--10
\put(05,33){\circle*{1.3}}

%=======================10-9
\put(05,33){\line(1,1){5}}

%=======================9-4
\put(10,38){\line(1,-4){5}}

%=======================9-7
\put(10,38){\line(1,-1){10}}

%--9
\put(10,38){\circle*{1.3}}

%--7
\put(20,28){\circle*{1.3}}

%%%%%%%%%%%%%%%%
\put(20,28){\line(-4,-3){20}}

%=======================7-4
\put(20,28){\line(-1,-2){5}}

%%%10:(5,38)
%=======================4-10
\put(15,18){\line(-2,3){10}}

%--4
\put(15,18){\circle*{1.3}}

%=======================4-3
\put(15,18){\line(-3,-1){15}}

%=======================4-2
\put(15,18){\line(-1,-2){5}}

%9:10,25
%=======================3-9
\put(00,13){\line(1,2){5}} \put(05,23){\line(1,3){5}}

%--3
\put(00,13){\circle*{1.3}}

%=======================2-3
\put(10,8){\line(-2,1){10}}

%--2
\put(10,8){\circle*{1.3}}

%--1 Additional
% \put(7.5,15.5){\circle*{1.3}}

%=======================1-5??
%\put(7.5,15.5){\line(1,4){2.5}}

%--8 Additional
%\put(10,25.5){\circle*{1.3}}

%%%%%%%%%%%%%%%%%%%%%%%%%%%%%%%%%%%%%%%%%%%%%%%%%%%%%%%%%%%%%%

%\put(27,41){\makebox(0,0)[bl]{Spanning tree 1}}

\put(29,42.6){\makebox(0,0)[bl]{Spanning}}
\put(31,39.6){\makebox(0,0)[bl]{tree \(T^{1}\)}}

%%%%%%%%%%%%%%%%%%%%%%%%%%%%%%%%%%%%%%%%%%%%%%%%%20
%--20
\put(25,28){\circle*{1.3}}

%=======================10-9
%\put(25,28){\line(1,1){5}}

\put(25,28){\line(0,-1){15}}

%\put(25,28){\line(3,-2){15}}

%%%%%%%%%%%%%%%%%%%%%%%%%%%%%%%%%%%%%%%%%%%%%%%%

%--10
\put(30,33){\circle*{1.3}}

%=======================10-9
\put(30,33){\line(1,1){5}}

%=======================9-4
\put(35,38){\line(1,-4){5}}

%=======================9-7
\put(35,38){\line(1,-1){10}}

%--9
\put(35,38){\circle*{1.3}}

%--7
\put(45,28){\circle*{1.3}}

%=======================7-4
%\put(45,28){\line(-1,-2){5}}

%%%10:(5,38)
%=======================4-10
%\put(40,18){\line(-2,3){5}}

%--4
\put(40,18){\circle*{1.3}}

%=======================4-3
\put(40,18){\line(-3,-1){15}}

%=======================4-2
\put(40,18){\line(-1,-2){5}}

%9:10,25
%=======================3-9
%\put(25,13){\line(1,2){5}} \put(30,23){\line(1,3){5}}

%--3
\put(25,13){\circle*{1.3}}

%=======================2-3
%\put(35,8){\line(-2,1){10}}

%--2
\put(35,8){\circle*{1.3}}

%--1 Additional
% \put(32.5,15.5){\circle*{1.3}}

%=======================1-5??
%\put(32.5,15.5){\line(1,4){2.5}}

%--8 Additional
%\put(35,25.5){\circle*{1.3}}

%%%%%%%%%%%%%%%%%%%%%%%%%%%%%%%%%%%%%%%%%%%%%%%%%%%%%%%%%%%%%%

\put(43,21){\makebox(0,0)[bl]{\(\Longrightarrow\)}}

%%%%%%%%%%%%%%%%%%%%%%%%%%%%%%%%%%%%%%%%%%%%%%%%%%%%%%%%%%%%%%

%%%%%%%%%%%%%%%%%%%%%%%%%%%%%%%%%%%%%%%%%%%%%%%%%%%%%%%%%%%%%%

%\put(47,22.6){\makebox(0,0)[bl]{\(\Longrightarrow\)}}

%%%%%%%%%%%%%%%%%%%%%%%%%%%%%%%%%%%%%%%%%%%%%%%%%%%%%%%%%%%%%%

\put(55,42.6){\makebox(0,0)[bl]{Spanning}}
\put(57,39.6){\makebox(0,0)[bl]{tree \(T^{*}\)}}

%%%%%%%%%%%%%%%%%%%%%%%%%%%%%%%%%%%%%%%%%%%%%%%%%
%--20
\put(50,28){\circle*{1.3}}

%=======================10-9
\put(50,28){\line(1,1){5}}

\put(50,28){\line(0,-1){15}}

%%%%%%%%%%%%%%%%%%%%%%%%%%%%%%%%%%%%%%%%%%%%%%%%

%--10
\put(55,33){\circle*{1.3}}

%=======================10-9
\put(55,33){\line(1,1){5}}

%--9
\put(60,38){\circle*{1.3}}

%--7
\put(70,28){\circle*{1.3}}

%=======================7-4
%\put(70,28){\line(-1,-2){5}}

\put(70,28){\line(-1,1){10}}

%%%10:(5,38)
%=======================4-10
\put(65,18){\line(-2,3){10}}

%--4
\put(65,18){\circle*{1.3}}

%=======================4-3
%%%%%\put(65,18){\line(-3,-1){15}}

%=======================4-2
\put(65,18){\line(-1,-2){5}}

%--3
\put(50,13){\circle*{1.3}}

%--2
\put(60,8){\circle*{1.3}}

%%%%%%%%%%%%%%%%%%%%%%%%%%%%%%%%%%%%%%%%%%%%%%%%%%%%%%%%%%%%%%
%%%%%%%%%%%%%%%%%%%%%%%%%%%%%%%%%%%%%%%%%%%%%%%%%%%%%%%%%%%%%%

\put(80,42.6){\makebox(0,0)[bl]{Spanning}}
\put(82,39.6){\makebox(0,0)[bl]{tree \(T^{2}\)}}

%%%%%%%%%%%%%%%%%%%%%%%%%%%%%%%%%%%%%%%%%%%%%%%%%
%--20
\put(75,28){\circle*{1.3}}

%=======================10-9
\put(75,28){\line(1,1){5}}

%\put(50,28){\line(0,-1){15}}

%\put(50,28){\line(3,-2){15}}

%%%%%%%%%%%%%%%%%%%%%%%%%%%%%%%%%%%%%%%%%%%%%%%%

%--10
\put(80,33){\circle*{1.3}}

%=======================10-9
\put(80,33){\line(1,1){5}}

%=======================9-4
%\put(60,38){\line(1,-4){5}}

%=======================9-7
%\put(60,38){\line(1,-1){10}}

%--9
\put(85,38){\circle*{1.3}}

%--7
\put(95,28){\circle*{1.3}}

%--7=> Steiner vertex
%\put(65,28){\circle*{1.3}} \put(65,28){\circle{2}}
%\put(65,28){\line(1,0){5}} \put(65,28){\line(0,-1){10}}
%
%\put(65,28){\line(-1,2){5}}

%=======================7-4
\put(95,28){\line(-1,-2){5}}

%%%10:(5,38)
%=======================4-10
\put(90,18){\line(-2,3){10}}

%--4
\put(90,18){\circle*{1.3}}

%=======================4-3
\put(90,18){\line(-3,-1){15}}

%=======================4-2
\put(90,18){\line(-1,-2){5}}

%9:10,25
%=======================3-9
%\put(50,13){\line(1,2){5}} \put(55,23){\line(1,3){5}}

%--3
\put(75,13){\circle*{1.3}}

%=======================2-3
%\put(60,8){\line(-2,1){10}}

%--2
\put(85,8){\circle*{1.3}}

%%%%%%%%%%%%%%%%%%%%%%%%%%%%%%%%%%%%%%%%%%%%%%%%%%%%%%%%%%%%%%

\end{picture}
\end{center}

 {\bf III.} Steiner trees (Fig. 7, set of possible Steiner vertices is \(Z = \{a,b,c,d\}\)):

 (i) \(S^{1} = (A^{1},E^{1})\), where ~\(A^{1}=A\bigcup Z^{1}\), ~\(Z^{1}=\{a,b\}\),

 \(E^{1}=\{(1,2),(1,a),(a,4),(a,6),(3,5),(b,5),(b,6),(b,7)\}\),

 (ii) \(S^{2} = (A^{2},E^{2})\), where ~\(A^{2}=A\bigcup Z^{2}\), ~\(Z^{2}=\{a,b,d\}\),

 \(E^{2}=\{(3,4),(1,d),(3,d),(a,d),(a,4),(a,6),(b,6),(b,5)),(b,7)\}\),

 (iii) \(S^{*} = (A^{*},E^{*})\), where where ~\(A^{*}=A\bigcup Z^{*}\), ~\(Z^{*}=\{a,c\}\),

 \(E^{*}=\{(1,2),(1,a),(a,4),(a,6),(c,3),(c,5),(c,6),(6,7)\}\).

\begin{center}
\begin{picture}(70,47)
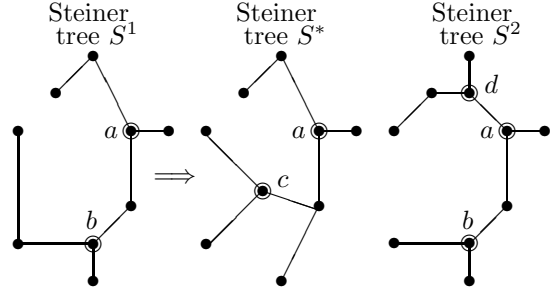
\put(07,00){\makebox(0,0)[bl]{Fig. 7. Restructuring of Steiner
tree}}

%\put(02,41){\makebox(0,0)[bl]{Initial graph}}

%>>>>>>>>>>>>>>>>>>>>>>>>>>>>>>>>>>>>>>>>>>>>>>> Numbers-labels

%\put(12,37){\makebox(0,0)[bl]{\(1\)}}
%\put(02,32){\makebox(0,0)[bl]{\(2\)}}
%\put(02,27){\makebox(0,0)[bl]{\(3\)}}
%\put(16,27){\makebox(0,0)[bl]{\(4\)}}
%\put(16.6,17){\makebox(0,0)[bl]{\(6\)}}
%\put(00,09){\makebox(0,0)[bl]{\(5\)}}
%\put(12,09){\makebox(0,0)[bl]{\(7\)}}

%%%%%%%%%%%%%%%%%%%%%%%%%%%%%%%%%%%%%%%%%%%%%%%%%
%%%%%%%%%%%%%%%%%%%%%%%%%%%%%%%%%%%%%%%%%%%%%%%%%%%%%%%%%%%%%%

%\put(27,41){\makebox(0,0)[bl]{Spanning tree}}

\put(04,42.6){\makebox(0,0)[bl]{Steiner}}
\put(05,39.6){\makebox(0,0)[bl]{tree \(S^{1}\)}}

%%%%%%%%%%%%%%%%%%%%%%%%%%%%%%%%%%%%%%%%%%%%%%%%%20
%--20
\put(00,28){\circle*{1.3}}

%=======================10-9
%\put(25,28){\line(1,1){5}}

\put(00,28){\line(0,-1){15}}

%\put(25,28){\line(3,-2){15}}

%%%%%%%%%%%%%%%%%%%%%%%%%%%%%%%%%%%%%%%%%%%%%%%%

%--10
\put(05,33){\circle*{1.3}}

%=======================10-9
\put(05,33){\line(1,1){5}}

%=======================9-4
%\put(35,38){\line(1,-4){5}}

%=======================9-7
%\put(35,38){\line(1,-1){10}}

%--9
\put(10,38){\circle*{1.3}}

%--7
\put(20,28){\circle*{1.3}}

%=======================7-4
%\put(45,28){\line(-1,-2){5}}

%%%10:(5,38)
%=======================4-10
%\put(40,18){\line(-2,3){5}}

%--4
\put(15,18){\circle*{1.3}}

%=======================4-3
%\put(40,18){\line(-3,-1){15}}

%=======================4-2
%\put(40,18){\line(-1,-2){5}}

%--7=> Steiner vertex
\put(15,28){\circle*{1.3}} \put(15,28){\circle{2}}
\put(15,28){\line(1,0){5}} \put(15,28){\line(0,-1){10}}

\put(15,28){\line(-1,2){5}}

\put(11.4,27){\makebox(0,0)[bl]{\(a\)}}

%9:10,25
%=======================3-9
%\put(25,13){\line(1,2){5}} \put(30,23){\line(1,3){5}}

%--3
\put(00,13){\circle*{1.3}}

%--7=> Steiner vertex
\put(10,13){\circle*{1.3}} \put(10,13){\circle{2}}
\put(10,13){\line(-1,0){10}} \put(10,13){\line(0,-1){5}}

\put(10,13){\line(1,1){5}}

\put(09,15){\makebox(0,0)[bl]{\(b\)}}

%=======================2-3
%\put(35,8){\line(-2,1){10}}

%--2
\put(10,8){\circle*{1.3}}

%--1 Additional
% \put(32.5,15.5){\circle*{1.3}}

%=======================1-5??
%\put(32.5,15.5){\line(1,4){2.5}}

%--8 Additional
%\put(35,25.5){\circle*{1.3}}

%%%%%%%%%%%%%%%%%%%%%%%%%%%%%%%%%%%%%%%%%%%%%%%%%%%%%%%%%%%%%%

\put(18,21){\makebox(0,0)[bl]{\(\Longrightarrow\)}}

%%%%%%%%%%%%%%%%%%%%%%%%%%%%%%%%%%%%%%%%%%%%%%%%%%%%%%%%%%%%%%
%%%%%%%%%%%%%%%%%%%%%%%%%%%%%%%%%%%%%%%%%%%%%%%%%%%%%%%%%%%%%%

%\put(27,41){\makebox(0,0)[bl]{Spanning tree}}

\put(29,42.6){\makebox(0,0)[bl]{Steiner}}
\put(30,39.6){\makebox(0,0)[bl]{tree \(S^{*}\)}}

%%%%%%%%%%%%%%%%%%%%%%%%%%%%%%%%%%%%%%%%%%%%%%%%%20
%--20
\put(25,28){\circle*{1.3}}

%=======================10-9
%\put(25,28){\line(1,1){5}}

%\put(25,28){\line(0,-1){15}}

%\put(25,28){\line(3,-2){15}}

%%%%%%%%%%%%%%%%%%%%%%%%%%%%%%%%%%%%%%%%%%%%%%%%

%--10
\put(30,33){\circle*{1.3}}

%=======================10-9
\put(30,33){\line(1,1){5}}

%=======================9-4
%\put(35,38){\line(1,-4){5}}

%=======================9-7
%\put(35,38){\line(1,-1){10}}

%--9
\put(35,38){\circle*{1.3}}

%--7
\put(45,28){\circle*{1.3}}

%=======================7-4
%\put(45,28){\line(-1,-2){5}}

%%%10:(5,38)
%=======================4-10
%\put(40,18){\line(-2,3){5}}

%--4
\put(40,18){\circle*{1.3}}

%=======================4-3
%\put(40,18){\line(-3,-1){15}}

%=======================4-2
%\put(40,18){\line(-1,-2){5}}

%--7=> Steiner vertex
\put(40,28){\circle*{1.3}} \put(40,28){\circle{2}}
\put(40,28){\line(1,0){5}} \put(40,28){\line(0,-1){10}}

\put(40,28){\line(-1,2){5}}

\put(36.4,27){\makebox(0,0)[bl]{\(a\)}}

%9:10,25
%=======================3-9
%\put(25,13){\line(1,2){5}} \put(30,23){\line(1,3){5}}

%--3
\put(25,13){\circle*{1.3}}

%%%%%%%%%%%%%%%%%%%%%%%%%%%%%%%%----------------------------------------------
%--7=> Steiner vertex
\put(32.5,20){\circle*{1.3}} \put(32.5,20){\circle{2}}

\put(32.5,20){\line(3,-1){7.5}} \put(32.5,20){\line(-1,-1){7.5}}
\put(32.5,20){\line(-1,1){7.5}}

\put(34.4,20.6){\makebox(0,0)[bl]{\(c\)}}

%=======================2-3
%\put(35,8){\line(-2,1){10}}

%%%%%%%%%%%%%%%%%%%%%%%%%%%%%%%%%%%%%%%%%%%%
%=======================4-3
%\put(40,18){\line(-3,-1){15}}

%=======================4-2
\put(40,18){\line(-1,-2){5}}

%%%%%%%%%%%%%%%%%%%%%%%%%%%%%%%%%%%%%%%%%%%%

%--2
\put(35,8){\circle*{1.3}}

%%%%%%%%%%%%%%%%%%%%%%%%%%%%%%%%%%%%%%%%%%%%%%%%%%%%%%%%%%%%%%

%\put(45,21){\makebox(0,0)[bl]{\(\Longrightarrow\)}}

%%%%%%%%%%%%%%%%%%%%%%%%%%%%%%%%%%%%%%%%%%%%%%%%%%%%%%%%%%%%%%

%\put(52,41.6){\makebox(0,0)[bl]{Steiner tree}}

\put(55,42.6){\makebox(0,0)[bl]{Steiner}}
\put(56,39.6){\makebox(0,0)[bl]{tree \(S^{2}\) }}

%%%%%%%%%%%%%%%%%%%%%%%%%%%%%%%%%%%%%%%%%%%%%%%%%
%--20
\put(50,28){\circle*{1.3}}

%=======================10-9
\put(50,28){\line(1,1){5}}

%\put(50,28){\line(0,-1){15}}

%\put(50,28){\line(3,-2){15}}

%%%%%%%%%%%%%%%%%%%%%%%%%%%%%%%%%%%%%%%%%%%%%%%%

%--10
\put(55,33){\circle*{1.3}}

%=======================10-9
%\put(55,33){\line(1,1){5}}

%=======================9-4
%\put(60,38){\line(1,-4){5}}

%=======================9-7
%\put(60,38){\line(1,-1){10}}

%--9
\put(60,38){\circle*{1.3}}

%--7=> Steiner vertex
\put(60,33){\circle*{1.3}} \put(60,33){\circle{2}}
\put(60,33){\line(0,1){5}} \put(60,33){\line(-1,0){5}}

\put(60,33){\line(1,-1){5}}

\put(62,33){\makebox(0,0)[bl]{\(d\)}}

%--7
\put(70,28){\circle*{1.3}}

%--7=> Steiner vertex
\put(65,28){\circle*{1.3}} \put(65,28){\circle{2}}
\put(65,28){\line(1,0){5}} \put(65,28){\line(0,-1){10}}

%\put(65,28){\line(-1,2){5}}

\put(61.4,27){\makebox(0,0)[bl]{\(a\)}}

%=======================7-4
%\put(70,28){\line(-1,-2){5}}

%%%10:(5,38)
%=======================4-10
%\put(65,18){\line(-2,3){10}}

%--4
\put(65,18){\circle*{1.3}}

%--3
\put(50,13){\circle*{1.3}}

%--7=> Steiner vertex
\put(60,13){\circle*{1.3}} \put(60,13){\circle{2}}
\put(60,13){\line(-1,0){10}} \put(60,13){\line(0,-1){5}}

\put(60,13){\line(1,1){5}}

\put(59,15){\makebox(0,0)[bl]{\(b\)}}

%--2
\put(60,8){\circle*{1.3}}

%%%%%%%%%%%%%%%%%%%%%%%%%%%%%%%%%%%%%%%%%%%%%%%%%%%%%%%%%%%%%%

\end{picture}
\end{center}

 Thus the restructuring problem for spanning tree is
% can be considered as the following
 (Fig. 6, a simple version):
  \[\min \rho ( T^{*} , T^{2})
  ~~~~ s.t. ~~~
   H(S^{1} \Rightarrow S^{*}) = ( \sum_{i \in  E^{1*-} } h^{-}_{i} +  \sum_{i \in E^{1*+}} h^{+}_{i} ~) \leq \widehat{h},\]
 where \(\widehat{h}\) is a constraint for the change cost,
 \(h^{-}(i)\) is a cost of deletion of element
 (i.e., edge) \(i \in E^{1}\), and
  \(h^{+}(i)\) is a cost of addition of element
  (i.e., edge) \(i  \in E \backslash  E^{1}\).

%
% Further,
 The restructuring problem for Steiner tree is
% can be considered as the following
 (Fig. 7, a simple version):
  \[\min \rho ( S^{*} , S^{2})\]
  \[s.t. ~~
   H(S^{1} \Rightarrow S^{*}) = ( \sum_{i \in  E^{1*-} } h^{-}_{i} +  \sum_{i \in E^{1*+}} h^{+}_{i}
   ~)  +
  ( \sum_{i \in  Z^{1*-} } w^{-}_{i} +  \sum_{i \in Z^{1*+}} w^{+}_{i} ~)
    \leq \widehat{h},\]
 where \(\widehat{h}\) is a constraint for the change cost,
 \(h^{-}(i)\) is a cost of deletion of element
 (i.e., edge) \(i \in E^{1}\),
  \(h^{+}(i)\) is a cost of addition of element
  (i.e., edge)
  \(i  \in  \widehat{E}^{*} \subseteq E \backslash  E^{1} \),
  \(w^{-}(j)\) is a cost of deletion of
  Steiner vertex \(j \in Z^{1}\),
  \(w^{+}(j)\) is a cost of addition of Steiner vertex
  \(j  \in \widehat{Z}^{*} \subseteq  Z \backslash Z^{1}\).

%%%%%%%%%%%%%%%%%%%%%%%%%%%%%%%%%%%%%%%%%%%%%%%%%%%%%%%%%%
%\section{Note on Solving Methods}

 In the main, the suggested restructuring problems
 are NP-hard and enumerative algorithms or heuristics can be used.
%
% On the other hand, it is possible to consider
% some directions for
 The design/selection of heuristics may be based  on some typical
 situations.
 First, the restructuring problems often are based
 on two selection subproblems:
 (a) deletion of elements and (b) addition of elements.
 This leads to possible usage of greedy-like algorithms.
 If the restructuring problem is based
 on exchange of elements (e.g., restructuring in assignment/allocation problem)
 local heuristics as k-exchange techniques can be used
 (e.g., 2-OPT, 3-OPT for travelling salesman problems).
 Further, methods of constraint programming can be widely used.
 Evidently, many well-known meta-heuristic methods can be used as
 well.
% (e.g., evolutionary optimization)
% (e.g., \cite{deb01}, \cite{osman96}).
%
 In addition,
 heuristic can be based on reducing of the basic restructuring
 problem, for example:
 (a) by problem type,
 (b) by problem dimension
 (e.g, selection of the most prospective change operations), etc.

\section{Illustrative Application Examples}

% In this section  two applied restructuring examples are considered.

%~~

  {\bf Example 1.} Reconfiguration of ``microelectronic components
  part''
  in wireless sensor (multiple choice problem)
%  \cite{levfim10}.
%
% The following simplified structure of system is examined:
%
% {\bf 1.}
% Microelectronic components part~
 ~\(M = R \star P \star D \star Q\) \cite{levfim10}:

    {\it 1.} Radio~  \(R\):~
    10 mw 916 MHz Radio~  \(R_{1}(3)\),
    1 mw 916 MHz Radio~  \(R_{2}(2)\),
    10 mw 600 MHz Radio~  \(R_{3}(2)\),
    1 mw 600 MHz Radio~ \(R_{4}(1)\).

    {\it 2.} Microprocessor~  \(P\):~
    MAXQ 2000~ \(P_{1}(1)\),
    AVR with embedded DAC/ ADC~  \(P_{2}(2)\),
    MSP~ \(P_{3}(3)\).

    {\it 3.} DAC/ADC~  \(D\):~
    Motorola~  \(D_{1}(2)\),
    AVR embedded DAC/ADC~  \(D_{2}(1)\),
    Analog Devices 1407~  \(D_{3}(2)\).

    {\it 4.} Memory~  \(Q\):~
     512 byte RAM~  \(Q_{1}(3)\),
     512 byte EEPROM~  \(Q_{2}(3)\),
     8 KByte Flash~ \(Q_{3}(2)\),
     1 MByte Flash~  \(Q_{4}(1)\).

\begin{center}
\begin{picture}(65,76)
%\begin{picture}(55,76)

\put(08,72){\makebox(0,0)[bl]{Table 1. Estimates of DAs}}

\put(00,0){\line(1,0){55}} \put(00,58){\line(1,0){55}}
\put(17,64){\line(1,0){38}} \put(00,70){\line(1,0){55}}

\put(00,0){\line(0,1){70}} \put(07,0){\line(0,1){70}}
\put(17,0){\line(0,1){70}} \put(37,0){\line(0,1){70}}
%\put(48,0){\line(0,1){70}}
\put(55,0){\line(0,1){70}}

\put(08.5,66){\makebox(0,0)[bl]{Cost}}
\put(08.5,60){\makebox(0,0)[bl]{(\(a_{ij}\))}}

\put(17.5,65.5){\makebox(0,0)[bl]{Change cost}}
\put(27,58){\line(0,1){6}}
\put(20,59){\makebox(0,0)[bl]{\(h_{ij}^{-}\)}}
\put(30,59){\makebox(0,0)[bl]{\(h_{ij}^{+}\)}}

\put(39,66){\makebox(0,0)[bl]{Priorities}}
\put(46,58){\line(0,1){6}}
\put(39,59){\makebox(0,0)[bl]{\(c_{ij}^{1}\)}}
\put(48,59){\makebox(0,0)[bl]{\(c_{ij}^{2}\)}}

%%%%%%%%%%%%%%%%%%%%%%%%%%%%%%%%%

\put(01,54){\makebox(0,0)[bl]{\(R_{1}\)}}
\put(01,50){\makebox(0,0)[bl]{\(R_{2}\)}}
\put(01,46){\makebox(0,0)[bl]{\(R_{3}\)}}
\put(01,42){\makebox(0,0)[bl]{\(R_{4}\)}}

\put(01,38){\makebox(0,0)[bl]{\(P_{1}\)}}
\put(01,34){\makebox(0,0)[bl]{\(P_{2}\)}}
\put(01,30){\makebox(0,0)[bl]{\(P_{3}\)}}

\put(01,26){\makebox(0,0)[bl]{\(D_{1}\)}}
\put(01,22){\makebox(0,0)[bl]{\(D_{2}\)}}
\put(01,18){\makebox(0,0)[bl]{\(D_{3}\)}}

\put(01,14){\makebox(0,0)[bl]{\(Q_{1}\)}}
\put(01,10){\makebox(0,0)[bl]{\(Q_{2}\)}}
\put(01,06){\makebox(0,0)[bl]{\(Q_{3}\)}}

\put(01,02){\makebox(0,0)[bl]{\(Q_{4}\)}}

%----------------------------R1

\put(11,54){\makebox(0,0)[bl]{\(6\)}}

\put(21,54){\makebox(0,0)[bl]{\(2\)}}
\put(31,54){\makebox(0,0)[bl]{\(2\)}}

\put(41,54){\makebox(0,0)[bl]{\(1\)}}
\put(49,54){\makebox(0,0)[bl]{\(1\)}}

%----------------------------R2

\put(11,50){\makebox(0,0)[bl]{\(5\)}}
\put(21,50){\makebox(0,0)[bl]{\(1\)}}
\put(31,50){\makebox(0,0)[bl]{\(1\)}}

\put(41,50){\makebox(0,0)[bl]{\(2\)}}
\put(49,50){\makebox(0,0)[bl]{\(3\)}}

%----------------------------R3

\put(11,46){\makebox(0,0)[bl]{\(3\)}}
\put(21,46){\makebox(0,0)[bl]{\(2\)}}
\put(31,46){\makebox(0,0)[bl]{\(1\)}}

\put(41,46){\makebox(0,0)[bl]{\(2\)}}
\put(49,46){\makebox(0,0)[bl]{\(1\)}}

%----------------------------R4

\put(11,42){\makebox(0,0)[bl]{\(2\)}}
\put(21,42){\makebox(0,0)[bl]{\(2\)}}
\put(31,42){\makebox(0,0)[bl]{\(2\)}}

\put(41,42){\makebox(0,0)[bl]{\(3\)}}
\put(49,42){\makebox(0,0)[bl]{\(2\)}}

%----------------------------P1

\put(11,38){\makebox(0,0)[bl]{\(5\)}}
\put(21,38){\makebox(0,0)[bl]{\(2\)}}
\put(31,38){\makebox(0,0)[bl]{\(3\)}}

\put(41,38){\makebox(0,0)[bl]{\(3\)}}
\put(49,38){\makebox(0,0)[bl]{\(2\)}}

%----------------------------P2

\put(10,34){\makebox(0,0)[bl]{\(10\)}}
\put(21,34){\makebox(0,0)[bl]{\(2\)}}
\put(31,34){\makebox(0,0)[bl]{\(2\)}}

\put(41,34){\makebox(0,0)[bl]{\(2\)}}
\put(49,34){\makebox(0,0)[bl]{\(3\)}}

%----------------------------P3

\put(10,30){\makebox(0,0)[bl]{\(30\)}}
\put(21,30){\makebox(0,0)[bl]{\(3\)}}
\put(31,30){\makebox(0,0)[bl]{\(2\)}}

\put(41,30){\makebox(0,0)[bl]{\(1\)}}
\put(49,30){\makebox(0,0)[bl]{\(2\)}}

%----------------------------D1

\put(11,26){\makebox(0,0)[bl]{\(2\)}}
\put(21,26){\makebox(0,0)[bl]{\(2\)}}
\put(31,26){\makebox(0,0)[bl]{\(3\)}}

\put(41,26){\makebox(0,0)[bl]{\(2\)}}
\put(49,26){\makebox(0,0)[bl]{\(3\)}}

%----------------------------D2

\put(11,22){\makebox(0,0)[bl]{\(1\)}}
\put(21,22){\makebox(0,0)[bl]{\(2\)}}
\put(31,22){\makebox(0,0)[bl]{\(2\)}}

\put(41,22){\makebox(0,0)[bl]{\(3\)}}
\put(49,22){\makebox(0,0)[bl]{\(2\)}}

%----------------------------D3

\put(11,18){\makebox(0,0)[bl]{\(2\)}}
\put(21,18){\makebox(0,0)[bl]{\(1\)}}
\put(31,18){\makebox(0,0)[bl]{\(1\)}}

\put(41,18){\makebox(0,0)[bl]{\(2\)}}
\put(49,18){\makebox(0,0)[bl]{\(1\)}}

%----------------------------Q1

\put(11,14){\makebox(0,0)[bl]{\(3\)}}
\put(21,14){\makebox(0,0)[bl]{\(2\)}}
\put(31,14){\makebox(0,0)[bl]{\(1\)}}

\put(41,14){\makebox(0,0)[bl]{\(1\)}}
\put(49,14){\makebox(0,0)[bl]{\(3\)}}

%----------------------------Q2

\put(11,10){\makebox(0,0)[bl]{\(2\)}}
\put(21,10){\makebox(0,0)[bl]{\(2\)}}
\put(31,10){\makebox(0,0)[bl]{\(2\)}}

\put(41,10){\makebox(0,0)[bl]{\(1\)}}
\put(49,10){\makebox(0,0)[bl]{\(3\)}}

%----------------------------Q3

\put(11,06){\makebox(0,0)[bl]{\(3\)}}
\put(21,06){\makebox(0,0)[bl]{\(1\)}}
\put(31,06){\makebox(0,0)[bl]{\(2\)}}

\put(41,06){\makebox(0,0)[bl]{\(2\)}}
\put(49,06){\makebox(0,0)[bl]{\(2\)}}

%----------------------------Q4

\put(11,02){\makebox(0,0)[bl]{\(3\)}}
\put(21,02){\makebox(0,0)[bl]{\(1\)}}
\put(31,02){\makebox(0,0)[bl]{\(1\)}}

\put(41,02){\makebox(0,0)[bl]{\(3\)}}
\put(49,02){\makebox(0,0)[bl]{\(2\)}}

\end{picture}
%\end{center}
%
%\begin{center}
%\begin{picture}(56,39)
\begin{picture}(50,73)

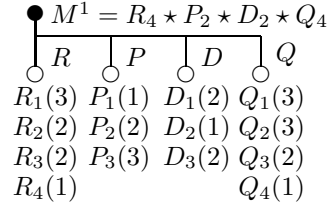
\put(03,40){\makebox(0,0)[bl]{Fig. 8. Structure of \(M^{1}\)}}

%>>>>>>>>>>>>>>>>>>>>>>>>>>>>>>>>>>>>>>>>>>>>>>>>>>>>>>
\put(01,57){\makebox(0,8)[bl]{\(R_{1}(3)\)}}
\put(01,53){\makebox(0,8)[bl]{\(R_{2}(2)\)}}
\put(01,49){\makebox(0,8)[bl]{\(R_{3}(2)\)}}
\put(01,45){\makebox(0,8)[bl]{\(R_{4}(1)\)}}

\put(11,57){\makebox(0,8)[bl]{\(P_{1}(1)\)}}
\put(11,53){\makebox(0,8)[bl]{\(P_{2}(2)\)}}
\put(11,49){\makebox(0,8)[bl]{\(P_{3}(3)\)}}
%\put(11,4){\makebox(0,8)[bl]{\(P_{8}(2)\)}}

\put(21,57){\makebox(0,8)[bl]{\(D_{1}(2)\)}}
\put(21,53){\makebox(0,8)[bl]{\(D_{2}(1)\)}}
\put(21,49){\makebox(0,8)[bl]{\(D_{3}(2)\)}}
%\put(21,4){\makebox(0,8)[bl]{\(R_{8}(1)\)}}

\put(31,57){\makebox(0,8)[bl]{\(Q_{1}(3)\)}}
\put(31,53){\makebox(0,8)[bl]{\(Q_{2}(3)\)}}
\put(31,49){\makebox(0,8)[bl]{\(Q_{3}(2)\)}}
\put(31,45){\makebox(0,8)[bl]{\(Q_{4}(1)\)}}

%--

\put(04,67){\line(1,0){30}}

\put(04,67){\line(0,-1){04}} \put(14,67){\line(0,-1){04}}
\put(24,67){\line(0,-1){04}} \put(34,67){\line(0,-1){04}}

%--

\put(04,62){\circle{2}} \put(14,62){\circle{2}}
\put(24,62){\circle{2}} \put(34,62){\circle{2}}

\put(06,63){\makebox(0,8)[bl]{\(R\) }}
\put(16,63){\makebox(0,8)[bl]{\(P\) }}
\put(26,63){\makebox(0,8)[bl]{\(D\) }}
\put(36,63){\makebox(0,8)[bl]{\(Q\) }}

%--

\put(04,67){\line(0,1){3}}

\put(04,70){\circle*{2}}

%\put(06,33){\makebox(0,8)[bl]{\(M = R \star P \star D \star Q\)}}

%\put(15,38){\makebox(0,8)[bl]{\(M^{1}_{1} = R_{4} \star P_{2}
%\star D_{2} \star Q_{4}\) }}

%\put(15,33){\makebox(0,8)[bl]{\(M^{3}_{2} = R_{4} \star P_{2}
%\star D_{2} \star Q_{4}\) }}

\put(06,68){\makebox(0,8)[bl]{\(M^{1} = R_{4} \star P_{2} \star
D_{2} \star Q_{4}\) }}

%\end{picture}
%\end{center}
%
%\begin{center}
%\begin{picture}(46,33)

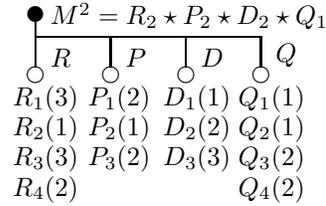
\put(03,0){\makebox(0,0)[bl]{Fig. 9. Structure of  \(M^{2}\)}}

%>>>>>>>>>>>>>>>>>>>>>>>>>>>>>>>>>>>>>>>>>>>>>>>>>>>>>>
\put(01,17){\makebox(0,8)[bl]{\(R_{1}(3)\)}}
\put(01,13){\makebox(0,8)[bl]{\(R_{2}(1)\)}}
\put(01,9){\makebox(0,8)[bl]{\(R_{3}(3)\)}}
\put(01,5){\makebox(0,8)[bl]{\(R_{4}(2)\)}}

\put(11,17){\makebox(0,8)[bl]{\(P_{1}(2)\)}}
\put(11,13){\makebox(0,8)[bl]{\(P_{2}(1)\)}}
\put(11,9){\makebox(0,8)[bl]{\(P_{3}(2)\)}}
%\put(11,4){\makebox(0,8)[bl]{\(P_{8}(2)\)}}

\put(21,17){\makebox(0,8)[bl]{\(D_{1}(1)\)}}
\put(21,13){\makebox(0,8)[bl]{\(D_{2}(2)\)}}
\put(21,9){\makebox(0,8)[bl]{\(D_{3}(3)\)}}
%\put(21,4){\makebox(0,8)[bl]{\(R_{8}(1)\)}}

\put(31,17){\makebox(0,8)[bl]{\(Q_{1}(1)\)}}
\put(31,13){\makebox(0,8)[bl]{\(Q_{2}(1)\)}}
\put(31,9){\makebox(0,8)[bl]{\(Q_{3}(2)\)}}
\put(31,5){\makebox(0,8)[bl]{\(Q_{4}(2)\)}}

%--

\put(04,27){\line(1,0){30}}

\put(04,27){\line(0,-1){04}} \put(14,27){\line(0,-1){04}}
\put(24,27){\line(0,-1){04}} \put(34,27){\line(0,-1){04}}

%--

\put(04,22){\circle{2}} \put(14,22){\circle{2}}
\put(24,22){\circle{2}} \put(34,22){\circle{2}}

\put(06,23){\makebox(0,8)[bl]{\(R\) }}
\put(16,23){\makebox(0,8)[bl]{\(P\) }}
\put(26,23){\makebox(0,8)[bl]{\(D\) }}
\put(36,23){\makebox(0,8)[bl]{\(Q\) }}

%--

\put(04,27){\line(0,1){3}}

\put(04,30){\circle*{2}}

%\put(06,33){\makebox(0,8)[bl]{\(M = R \star P \star D \star Q\)}}

%\put(15,38){\makebox(0,8)[bl]{\(M^{3}_{1} = R_{4} \star P_{2}
%\star D_{2} \star Q_{4}\) }}

%\put(15,33){\makebox(0,8)[bl]{\(M^{3}_{2} = R_{4} \star P_{2}
%\star D_{2} \star Q_{4}\) }}

\put(06,28){\makebox(0,8)[bl]{\(M^{2} = R_{2} \star P_{2} \star
D_{2} \star Q_{1}\) }}

\end{picture}
\end{center}

% In \cite{levfim10} the design problem involves multicriteria description of design elements (i.e., design alternatives DAs)
% that is based on criteria.

%
 Here it is assumed that solutions
% \(M^{1}\) and \(M^{2}\)
 are based on multiple choice problem
 (in \cite{levfim10} the solving process was based on morphological clique problem
 while taking into account compatibility of selected DAs).
 Thus two solutions \(M^{1}\) (for \(t=\tau_{1}\), Fig. 8) and
 \(M^{2}\) (for \(t=\tau_{2}\), Fig. 9) are examined
 (in \cite{levfim10} the solutions correspond to trajectory
 design: stage 1 and stage 3).
 Table 1 contains estimates of DAs (expert judgment).
 Estimates of cost (Table 1) and priorities (Fig. 8, Fig. 9, in parentheses)
 correspond to examples in \cite{levfim10}.
 Here \(c_{ij} = 4-p_{ij}\).
%
%
% Now let us consider a corresponding restructuring
% \(M^{1}\) to search for \(M^{*}\).
 Two possible change operations can be considered (\( M^{1} \Rightarrow M^{*}\), \(M^{*}\) is close to \(M^{2}\) ):

 (a) \(R_{4} \rightarrow R_{2}  \),
 \(h_{a}^{-} = 2\), \(h_{a}^{+} = 1\)
 (corresponding Boolean variable \(x_{a} \in \{0,1\}\)),
%  and

 (b) \(Q_{4} \rightarrow Q_{1}  \),
 \(h_{b}^{-} = 1\), \(h_{b}^{+} = 1\)
 (corresponding Boolean variable \(x_{b} \in \{0,1\}\)).

 As a result, the following simplified knapsack problem can be used:
%  (i.e., selection of change operations):

   \[  \max ~( ~(c^{2}(R_{2}) - c^{2}(R_{4}) )~ x_{a} +  (c^{2}(Q_{1}) - c^{2}(Q_{4}) ) ~x_{b}    ~ ) \]
   \[s.t. ~~~ H (M^{*} \rightarrow M^{2}) ~ = ~ (~h^{-} (R_{4} \rightarrow R_{2}) + h^{+} (R_{4} \rightarrow R_{2} )~) ~x_{a}
   +\]
   \[ (h^{-}(Q_{4} \rightarrow Q_{1} )  + h^{+}(Q_{4} \rightarrow Q_{1} )~)~ x_{b}  \leq
   \widehat{h}.
   \]

 Finally, the restructuring solutions are:
 ~(i) \(\widehat{h} = 2\):
 ~\(M^{*1} = R_{4} \star P_{2} \star D_{2} \star Q_{1}\),
 ~(ii) \(\widehat{h} = 3\):
 ~\(M^{*2} = R_{2} \star P_{2} \star D_{2} \star Q_{4}\),
 ~(iii) \(\widehat{h} = 5\):
 ~\(M^{*3} = M^{2} = R_{2} \star P_{2} \star D_{2} \star Q_{1}\).
 Evidently, real restructuring problems can be more complicated.

~~

 {\bf Example 2.} Reassignment of users to access points
% (assignment problem)
 (\cite{lev10a}, \cite{levpet10}).
 Here the initial multicriteria assignment problem involves
 21 users and 6 access points.
% Tables 2, 3 4
% (a simplified version from
% (\cite{lev10a}).
%
%
 Tables 2, 4 contain
 some parameters for users (\(A\))
 (coordinates (\(x_{i},y_{i},z_{i}\)),
  required frequency spectrum \(f_{j}\),
%  required level of security \(p_{j}\),
  required level of reliability \(r_{j}\), etc.)
 and
 some parameters for 6 access points (\(B = \{j\} =\{1,2,3,4,5,6\}\))
 (coordinates (\(x_{j},y_{j},z_{j}\)),
  frequency spectrum \(f_{j}\),
%  maximal
  number of connections \(n_{j}\),
  level of reliability \(r_{j}\))
  (\cite{lev10a}, \cite{levpet10}).
 A simplified version of assignment problem from \cite{lev10a} is considered.
 Two regions are examined:
 an initial region and an additional region (Fig. 10).
 In \cite{lev10a} the problem was solved for two cases:
 (i) separated assignment \(S^{1}\) (Fig. 10),
 (ii) joint assignment \(S^{2}\) (Fig. 11).
 The restructured problem is considered
 as a modification (change) of \(S^{1}\) into \(S^{*}\).
% (\(S^{*}\)  \(S^{2}\))
%
 To reduce the problem
 it is reasonable
 the select a subset of users
 (a ``change zone'' near borders between regions):
 ~\(\widetilde{A} = \{i\} =\{3,5,8,12,13,14,17,19,21\}\).
 Thus,
 it is necessary to assign each element of
 \(\widetilde{A}\) into an access point of ~\(B\).

\begin{center}
%\begin{picture}(42,38)
\begin{picture}(44,27)

\put(00,34){\makebox(0,0)[bl]{Table 2. Access points}}

\put(00,00){\line(1,0){42}} \put(00,26){\line(1,0){42}}
\put(00,32){\line(1,0){42}}

\put(00,00){\line(0,1){32}} \put(04,00){\line(0,1){32}}
\put(11,00){\line(0,1){32}} \put(18,00){\line(0,1){32}}
\put(24,00){\line(0,1){32}} \put(30,00){\line(0,1){32}}
\put(36,00){\line(0,1){32}} \put(42,00){\line(0,1){32}}
%\put(48,00){\line(0,1){21}}

\put(01,28){\makebox(0,0)[bl]{\(j\)}}
\put(05.5,28){\makebox(0,0)[bl]{\(x_{j}\)}}
\put(12.5,28){\makebox(0,0)[bl]{\(y_{j}\)}}
\put(19,28){\makebox(0,0)[bl]{\(z_{j}\)}}
\put(25,28){\makebox(0,0)[bl]{\(f_{j}\)}}
\put(31,28){\makebox(0,0)[bl]{\(n_{j}\)}}
\put(37,28){\makebox(0,0)[bl]{\(r_{j}\)}}
%\put(43,16){\makebox(0,0)[bl]{\(d_{j}\)}}

%--

\put(01,22){\makebox(0,0)[bl]{\(1\)}}
\put(05.5,22){\makebox(0,0)[bl]{\(50\)}}
\put(11.5,22){\makebox(0,0)[bl]{\(157\)}}
\put(19,22){\makebox(0,0)[bl]{\(10\)}}
\put(25,22){\makebox(0,0)[bl]{\(30\)}}
\put(32,22){\makebox(0,0)[bl]{\(4\)}}
\put(37,22){\makebox(0,0)[bl]{\(10\)}}
%\put(43,10){\makebox(0,0)[bl]{\(10\)}}

%--

\put(01,18){\makebox(0,0)[bl]{\(2\)}}
\put(05.5,18){\makebox(0,0)[bl]{\(72\)}}
\put(11.5,18){\makebox(0,0)[bl]{\(102\)}}
\put(19,18){\makebox(0,0)[bl]{\(10\)}}
\put(25,18){\makebox(0,0)[bl]{\(42\)}}
\put(32,18){\makebox(0,0)[bl]{\(6\)}}
\put(37,18){\makebox(0,0)[bl]{\(10\)}}
%\put(44,06){\makebox(0,0)[bl]{\(9\)}}

%--=================================================

\put(01,14){\makebox(0,0)[bl]{\(3\)}}
\put(05.5,14){\makebox(0,0)[bl]{\(45\)}}
\put(12.5,14){\makebox(0,0)[bl]{\(52\)}}
\put(19,14){\makebox(0,0)[bl]{\(10\)}}
\put(25,14){\makebox(0,0)[bl]{\(45\)}}
\put(31,14){\makebox(0,0)[bl]{\(10\)}}
\put(37,14){\makebox(0,0)[bl]{\(10\)}}
%\put(43,02){\makebox(0,0)[bl]{\(10\)}}

%--

\put(01,10){\makebox(0,0)[bl]{\(4\)}}
\put(04.5,10){\makebox(0,0)[bl]{\(150\)}}
\put(11.5,10){\makebox(0,0)[bl]{\(165\)}}
\put(19,10){\makebox(0,0)[bl]{\(10\)}}
\put(25,10){\makebox(0,0)[bl]{\(30\)}}
\put(32,10){\makebox(0,0)[bl]{\(5\)}}
\put(37,10){\makebox(0,0)[bl]{\(15\)}}
%\put(44,10){\makebox(0,0)[bl]{\(8\)}}

%--

\put(01,06){\makebox(0,0)[bl]{\(5\)}}
\put(04.5,06){\makebox(0,0)[bl]{\(140\)}}
\put(11.5,06){\makebox(0,0)[bl]{\(112\)}}
\put(19,06){\makebox(0,0)[bl]{\(10\)}}
\put(25,06){\makebox(0,0)[bl]{\(32\)}}
\put(32,06){\makebox(0,0)[bl]{\(5\)}}
\put(38,06){\makebox(0,0)[bl]{\(8\)}}
%\put(44,06){\makebox(0,0)[bl]{\(8\)}}

%--=================================================

\put(01,02){\makebox(0,0)[bl]{\(6\)}}
\put(04.5,02){\makebox(0,0)[bl]{\(147\)}}
\put(12.5,02){\makebox(0,0)[bl]{\(47\)}}
\put(19,02){\makebox(0,0)[bl]{\(10\)}}
\put(25,02){\makebox(0,0)[bl]{\(30\)}}
\put(32,02){\makebox(0,0)[bl]{\(5\)}}
\put(37,02){\makebox(0,0)[bl]{\(15\)}}
%\put(44,02){\makebox(0,0)[bl]{\(7\)}}
%------------------------------------------------

\end{picture}
%\end{center}
%
%
%\begin{center}
\begin{picture}(65,56)
%\begin{picture}(50,27)

\put(10,52){\makebox(0,0)[bl]{Table 3. Users-access points}}

\put(00,00){\line(1,0){65}} \put(00,38){\line(1,0){65}}
\put(05,44){\line(1,0){60}}

\put(00,50){\line(1,0){65}}

\put(00,00){\line(0,1){50}} \put(05,00){\line(0,1){50}}
\put(15,00){\line(0,1){44}} \put(25,00){\line(0,1){44}}
\put(35,00){\line(0,1){44}} \put(45,00){\line(0,1){44}}
\put(55,00){\line(0,1){44}} \put(65,00){\line(0,1){50}}
%\put(48,00){\line(0,1){21}}

\put(11,45.1){\makebox(0,0)[bl]{Access points \(\{j\}\):~
\(h_{i,j}^{-}\),\(h_{i,j}^{+}\),\(c_{i,j}\)}}

\put(1.8,46){\makebox(0,0)[bl]{\(i\)}}
\put(09,40){\makebox(0,0)[bl]{\(1\)}}
\put(19,40){\makebox(0,0)[bl]{\(2\)}}
\put(29,40){\makebox(0,0)[bl]{\(3\)}}
\put(39,40){\makebox(0,0)[bl]{\(4\)}}
\put(49,40){\makebox(0,0)[bl]{\(5\)}}
\put(59,40){\makebox(0,0)[bl]{\(6\)}}
%\put(43,16){\makebox(0,0)[bl]{\(d_{j}\)}}

%--

\put(1.4,34){\makebox(0,0)[bl]{\(3\)}}

\put(6,33.5){\makebox(0,0)[bl]{\(3,2,2\)}}
\put(16,33.5){\makebox(0,0)[bl]{\(2,1,3\)}}
\put(26,33.5){\makebox(0,0)[bl]{\(1,0,3\)}}
\put(36,33.5){\makebox(0,0)[bl]{\(3,1,3\)}}
\put(46,33.5){\makebox(0,0)[bl]{\(2,1,0\)}}
\put(56,33.5){\makebox(0,0)[bl]{\(1,1,0\)}}
%\put(43,10){\makebox(0,0)[bl]{\(10\)}}

%--

\put(1.4,30){\makebox(0,0)[bl]{\(5\)}}

\put(6,29.5){\makebox(0,0)[bl]{\(2,1,1\)}}
\put(16,29.5){\makebox(0,0)[bl]{\(1,3,1\)}}
\put(26,29.5){\makebox(0,0)[bl]{\(1,2,1\)}}
\put(36,29.5){\makebox(0,0)[bl]{\(3,2,1\)}}
\put(46,29.5){\makebox(0,0)[bl]{\(1,1,1\)}}
\put(56,29.5){\makebox(0,0)[bl]{\(1,1,1\)}}
%\put(43,10){\makebox(0,0)[bl]{\(10\)}}

%--

\put(1.4,26){\makebox(0,0)[bl]{\(8\)}}

\put(06,25.5){\makebox(0,0)[bl]{\(1,1,3\)}}
\put(16,25.5){\makebox(0,0)[bl]{\(1,1,3\)}}
\put(26,25.5){\makebox(0,0)[bl]{\(1,1,3\)}}
\put(36,25.5){\makebox(0,0)[bl]{\(1,1,0\)}}
\put(46,25.5){\makebox(0,0)[bl]{\(1,1,3\)}}
\put(56,25.5){\makebox(0,0)[bl]{\(2,2,2\)}}
%\put(43,10){\makebox(0,0)[bl]{\(10\)}}

%--

\put(0.4,22){\makebox(0,0)[bl]{\(12\)}}

\put(06,21.5){\makebox(0,0)[bl]{\(2,2,3\)}}
\put(16,21.5){\makebox(0,0)[bl]{\(1,2,3\)}}
\put(26,21.5){\makebox(0,0)[bl]{\(1,2,3\)}}
\put(36,21.5){\makebox(0,0)[bl]{\(3,1,0\)}}
\put(46,21.5){\makebox(0,0)[bl]{\(2,1,0\)}}
\put(56,21.5){\makebox(0,0)[bl]{\(1,1,0\)}}
%\put(43,10){\makebox(0,0)[bl]{\(10\)}}

%--

\put(0.4,18){\makebox(0,0)[bl]{\(13\)}}

\put(06,17.5){\makebox(0,0)[bl]{\(1,1,3\)}}
\put(16,17.5){\makebox(0,0)[bl]{\(1,1,3\)}}
\put(26,17.5){\makebox(0,0)[bl]{\(1,1,3\)}}
\put(36,17.5){\makebox(0,0)[bl]{\(2,1,0\)}}
\put(46,17.5){\makebox(0,0)[bl]{\(2,2,1\)}}
\put(56,17.5){\makebox(0,0)[bl]{\(1,1,3\)}}
%\put(44,06){\makebox(0,0)[bl]{\(9\)}}

%--=================================================

\put(0.4,14){\makebox(0,0)[bl]{\(14\)}}

\put(06,13.5){\makebox(0,0)[bl]{\(1,1,1\)}}
\put(16,13.5){\makebox(0,0)[bl]{\(2,2,2\)}}
\put(26,13.5){\makebox(0,0)[bl]{\(1,2,0\)}}
\put(36,13.5){\makebox(0,0)[bl]{\(1,1,1\)}}
\put(46,13.5){\makebox(0,0)[bl]{\(1,1,1\)}}
\put(56,13.5){\makebox(0,0)[bl]{\(1,1,0\)}}
%\put(43,02){\makebox(0,0)[bl]{\(10\)}}

%--

\put(0.4,10){\makebox(0,0)[bl]{\(17\)}}

\put(06,09.5){\makebox(0,0)[bl]{\(1,1,2\)}}
\put(16,09.5){\makebox(0,0)[bl]{\(1,1,1\)}}
\put(26,09.5){\makebox(0,0)[bl]{\(1,0,1\)}}
\put(36,09.5){\makebox(0,0)[bl]{\(3,1,1\)}}
\put(46,09.5){\makebox(0,0)[bl]{\(1,1,1\)}}
\put(56,09.5){\makebox(0,0)[bl]{\(1,1,1\)}}
%\put(44,10){\makebox(0,0)[bl]{\(8\)}}

%--

\put(0.4,06){\makebox(0,0)[bl]{\(19\)}}

\put(06,05.5){\makebox(0,0)[bl]{\(1,1,0\)}}
\put(16,05.5){\makebox(0,0)[bl]{\(1,1,3\)}}
\put(26,05.5){\makebox(0,0)[bl]{\(1,2,3\)}}
\put(36,05.5){\makebox(0,0)[bl]{\(3,2,0\)}}
\put(46,05.5){\makebox(0,0)[bl]{\(1,1,3\)}}
\put(56,05.5){\makebox(0,0)[bl]{\(1,1,2\)}}
%\put(44,06){\makebox(0,0)[bl]{\(8\)}}

%--=================================================

\put(0.4,02){\makebox(0,0)[bl]{\(21\)}}

\put(06,01.5){\makebox(0,0)[bl]{\(1,1,0\)}}
\put(16,01.5){\makebox(0,0)[bl]{\(1,2,3\)}}
\put(26,01.5){\makebox(0,0)[bl]{\(1,1,2\)}}
\put(36,01.5){\makebox(0,0)[bl]{\(3,1,1\)}}
\put(46,01.5){\makebox(0,0)[bl]{\(1,1,1\)}}
\put(56,01.5){\makebox(0,0)[bl]{\(1,1,1\)}}
%\put(44,02){\makebox(0,0)[bl]{\(7\)}}
%------------------------------------------------

\end{picture}
\end{center}

 The considered simplified restructuring problem
 is based on set of change operations:
 (1) user 3, change of connection: ~\(1\rightarrow 4\)~
% (corresponding
 (Boolean variable \(x_{1}\)),
 (2) user 13, change of connection: ~\(3\rightarrow 6\)~
%  (corresponding
 (Boolean variable \(x_{2}\)),
 (3) user 21, change of connection: ~\(5\rightarrow 2\)~
%  (corresponding
  (Boolean variable \(x_{3}\)).
 Table 3 contains estimates of change costs (expert judgment)
 and ``integrated profits'' of correspondence between users and access
 points
 from  (\cite{lev10a}, \cite{levpet10}).
 The
% restructuring
 problem is:
 \[ \max ~(~ c_{3,4}~x_{1}+c_{13,6}~x_{2}+c_{21,2}~x_{3}~)\]
 \[s.t.~~ (~
 (h^{-}_{3,1}+h^{+}_{3,4}) ~x_{1} +
 (h^{-}_{13,3}+h^{+}_{13,6}) ~x_{2} +
 (h^{-}_{21,51}+h^{+}_{21,2}) ~x_{3}~)
 \leq \widehat{h}. \]
 The reassignment
 ~\( S^{*} \)~ is depicted in Fig. 12
 (i.e., \(x_{1}=0, x_{1}=1, x_{3}=1\), \(\widehat{h}=5\)).

\begin{center}
%\begin{picture}(48,99)
\begin{picture}(54,99)

\put(09,95){\makebox(0,0)[bl]{Table 4. Users}}

\put(00,00){\line(1,0){41}} \put(00,86){\line(1,0){41}}
\put(00,93){\line(1,0){41}}

\put(00,00){\line(0,1){93}} \put(06,00){\line(0,1){93}}
\put(13,00){\line(0,1){93}} \put(20,00){\line(0,1){93}}
\put(27,00){\line(0,1){93}} \put(34,00){\line(0,1){93}}
\put(41,00){\line(0,1){93}}

%\put(48,00){\line(0,1){93}}
%\put(55,00){\line(0,1){65}}

%\put(62,00){\line(0,1){65}}

\put(02,88){\makebox(0,0)[bl]{\(i\)}}

\put(08,88){\makebox(0,0)[bl]{\(x_{i}\)}}
\put(15,88){\makebox(0,0)[bl]{\(y_{i}\)}}
\put(22,88){\makebox(0,0)[bl]{\(z_{i}\)}}

\put(29,88){\makebox(0,0)[bl]{\(f_{i}\)}}
%\put(36,60){\makebox(0,0)[bl]{\(k_{i}\)}}

%\put(36,88){\makebox(0,0)[bl]{\(p_{i}\)}}
\put(36,88){\makebox(0,0)[bl]{\(r_{i}\)}}
%\put(57,60){\makebox(0,0)[bl]{\(d_{i}\)}}

%----------------------------------------
%%98 - 54 = 44
%--1
\put(02,82){\makebox(0,0)[bl]{\(1\)}}
\put(08,82){\makebox(0,0)[bl]{\(30\)}}
\put(14,82){\makebox(0,0)[bl]{\(165\)}}
\put(22.5,82){\makebox(0,0)[bl]{\(5\)}}

\put(28.5,82){\makebox(0,0)[bl]{\(10\)}}

%\put(36.5,80){\makebox(0,0)[bl]{\(2\)}}

%\put(36.5,82){\makebox(0,0)[bl]{\(2\)}}
\put(36.5,82){\makebox(0,0)[bl]{\(5\)}}
%\put(57.5,80){\makebox(0,0)[bl]{\(8\)}}

%--2
\put(02,78){\makebox(0,0)[bl]{\(2\)}}
\put(08,78){\makebox(0,0)[bl]{\(58\)}}
\put(14,78){\makebox(0,0)[bl]{\(174\)}}
\put(22.5,78){\makebox(0,0)[bl]{\(5\)}}

\put(29.5,78){\makebox(0,0)[bl]{\(5\)}}

%\put(36.5,76){\makebox(0,0)[bl]{\(1\)}}

%\put(36.5,78){\makebox(0,0)[bl]{\(1\)}}
\put(36.5,78){\makebox(0,0)[bl]{\(9\)}}
%\put(57.5,76){\makebox(0,0)[bl]{\(6\)}}

%--3
\put(02,74){\makebox(0,0)[bl]{\(3\)}}
\put(08,74){\makebox(0,0)[bl]{\(95\)}}
\put(14,74){\makebox(0,0)[bl]{\(156\)}}
\put(22.5,74){\makebox(0,0)[bl]{\(0\)}}
\put(29.5,74){\makebox(0,0)[bl]{\(6\)}}

%\put(36.5,72){\makebox(0,0)[bl]{\(1\)}}

%\put(36.5,74){\makebox(0,0)[bl]{\(1\)}}
\put(36.5,74){\makebox(0,0)[bl]{\(6\)}}
%\put(57.5,72){\makebox(0,0)[bl]{\(8\)}}

%--8->7-4
\put(02,70){\makebox(0,0)[bl]{\(4\)}}
\put(08,70){\makebox(0,0)[bl]{\(52\)}}
\put(14,70){\makebox(0,0)[bl]{\(134\)}}
\put(22.5,70){\makebox(0,0)[bl]{\(5\)}}
\put(29.5,70){\makebox(0,0)[bl]{\(6\)}}

%\put(36.5,68){\makebox(0,0)[bl]{\(1\)}}

%\put(36.5,70){\makebox(0,0)[bl]{\(1\)}}
\put(36.5,70){\makebox(0,0)[bl]{\(8\)}}
%\put(57.5,68){\makebox(0,0)[bl]{\(7\)}}

%--9->8
\put(02,66){\makebox(0,0)[bl]{\(5\)}}
\put(08,66){\makebox(0,0)[bl]{\(85\)}}
\put(14,66){\makebox(0,0)[bl]{\(134\)}}
\put(22.5,66){\makebox(0,0)[bl]{\(3\)}}
\put(29.5,66){\makebox(0,0)[bl]{\(6\)}}

%\put(36.5,64){\makebox(0,0)[bl]{\(1\)}}

%\put(36.5,66){\makebox(0,0)[bl]{\(1\)}}
\put(36.5,66){\makebox(0,0)[bl]{\(7\)}}
%\put(57.5,64){\makebox(0,0)[bl]{\(7\)}}

%--13->12
\put(02,62){\makebox(0,0)[bl]{\(6\)}}
\put(08,62){\makebox(0,0)[bl]{\(27\)}}
\put(14,62){\makebox(0,0)[bl]{\(109\)}}
\put(22.5,62){\makebox(0,0)[bl]{\(7\)}}
\put(29.5,62){\makebox(0,0)[bl]{\(8\)}}

%\put(36.5,60){\makebox(0,0)[bl]{\(1\)}}

%\put(36.5,62){\makebox(0,0)[bl]{\(3\)}}
\put(36.5,62){\makebox(0,0)[bl]{\(5\)}}
%\put(57.5,60){\makebox(0,0)[bl]{\(8\)}}

%--14->13
\put(02,58){\makebox(0,0)[bl]{\(7\)}}
\put(08,58){\makebox(0,0)[bl]{\(55\)}}
\put(14,58){\makebox(0,0)[bl]{\(105\)}}
\put(22.5,58){\makebox(0,0)[bl]{\(2\)}}
\put(29.5,58){\makebox(0,0)[bl]{\(7\)}}

%\put(36.5,50){\makebox(0,0)[bl]{\(1\)}}

%\put(36.5,58){\makebox(0,0)[bl]{\(2\)}}
\put(35.5,58){\makebox(0,0)[bl]{\(10\)}}
%\put(57.5,50){\makebox(0,0)[bl]{\(6\)}}

%--15->14
\put(02,54){\makebox(0,0)[bl]{\(8\)}}
\put(08,54){\makebox(0,0)[bl]{\(98\)}}
\put(15,54){\makebox(0,0)[bl]{\(89\)}}
\put(22.5,54){\makebox(0,0)[bl]{\(3\)}}
\put(28.5,54){\makebox(0,0)[bl]{\(10\)}}

%\put(36.5,56){\makebox(0,0)[bl]{\(3\)}}

%\put(36.5,54){\makebox(0,0)[bl]{\(1\)}}
\put(35.5,54){\makebox(0,0)[bl]{\(10\)}}
%\put(57.5,56){\makebox(0,0)[bl]{\(7\)}}

%--20->16
\put(02,50){\makebox(0,0)[bl]{\(9\)}}
\put(08,50){\makebox(0,0)[bl]{\(25\)}}
\put(15,50){\makebox(0,0)[bl]{\(65\)}}
\put(22.5,50){\makebox(0,0)[bl]{\(2\)}}
\put(29.5,50){\makebox(0,0)[bl]{\(7\)}}

%\put(36.5,48){\makebox(0,0)[bl]{\(1\)}}

%\put(36.5,50){\makebox(0,0)[bl]{\(3\)}}
\put(36.5,50){\makebox(0,0)[bl]{\(5\)}}
%\put(57.5,48){\makebox(0,0)[bl]{\(6\)}}

%--21->17
\put(01,46){\makebox(0,0)[bl]{\(10\)}}
\put(08,46){\makebox(0,0)[bl]{\(52\)}}
\put(15,46){\makebox(0,0)[bl]{\(81\)}}
\put(22.5,46){\makebox(0,0)[bl]{\(1\)}}
\put(28.5,46){\makebox(0,0)[bl]{\(10\)}}

%\put(36.5,44){\makebox(0,0)[bl]{\(3\)}}

%\put(36.5,46){\makebox(0,0)[bl]{\(1\)}}
\put(36.5,46){\makebox(0,0)[bl]{\(8\)}}
%\put(56.5,44){\makebox(0,0)[bl]{\(10\)}}

%--28->20
\put(01,42){\makebox(0,0)[bl]{\(11\)}}
\put(08,42){\makebox(0,0)[bl]{\(65\)}}
\put(15,42){\makebox(0,0)[bl]{\(25\)}}
\put(22.5,42){\makebox(0,0)[bl]{\(7\)}}
\put(29.5,42){\makebox(0,0)[bl]{\(6\)}}

%\put(36.5,40){\makebox(0,0)[bl]{\(1\)}}

%\put(36.5,42){\makebox(0,0)[bl]{\(2\)}}
\put(36.5,42){\makebox(0,0)[bl]{\(9\)}}
%\put(57.5,40){\makebox(0,0)[bl]{\(8\)}}

%--29->21
\put(01,38){\makebox(0,0)[bl]{\(12\)}}
\put(08,38){\makebox(0,0)[bl]{\(93\)}}
\put(15,38){\makebox(0,0)[bl]{\(39\)}}
\put(22.5,38){\makebox(0,0)[bl]{\(1\)}}
\put(28.5,38){\makebox(0,0)[bl]{\(10\)}}

%\put(36.5,36){\makebox(0,0)[bl]{\(2\)}}

%\put(36.5,38){\makebox(0,0)[bl]{\(1\)}}
\put(35.5,38){\makebox(0,0)[bl]{\(10\)}}
%\put(57.5,36){\makebox(0,0)[bl]{\(9\)}}

%--new 23
\put(01,34){\makebox(0,0)[bl]{\(13\)}}
\put(07,34){\makebox(0,0)[bl]{\(172\)}}
\put(15,34){\makebox(0,0)[bl]{\(26\)}}
\put(22.5,34){\makebox(0,0)[bl]{\(2\)}}
\put(28.5,34){\makebox(0,0)[bl]{\(10\)}}

%\put(36.5,32){\makebox(0,0)[bl]{\(1\)}}

%\put(36.5,34){\makebox(0,0)[bl]{\(2\)}}
\put(36.5,34){\makebox(0,0)[bl]{\(7\)}}
%\put(57.5,32){\makebox(0,0)[bl]{\(6\)}}

%----------------------------------------
%----------------------------------------
%----------------------------------------
%--4
\put(01,30){\makebox(0,0)[bl]{\(14\)}}
\put(07,30){\makebox(0,0)[bl]{\(110\)}}
\put(14,30){\makebox(0,0)[bl]{\(169\)}}
\put(22.5,30){\makebox(0,0)[bl]{\(5\)}}
\put(29.5,30){\makebox(0,0)[bl]{\(7\)}}

%\put(36.5,28){\makebox(0,0)[bl]{\(1\)}}

%\put(36.5,30){\makebox(0,0)[bl]{\(2\)}}
\put(36.5,30){\makebox(0,0)[bl]{\(5\)}}
%\put(57.5,28){\makebox(0,0)[bl]{\(6\)}}

%--5
\put(01,26){\makebox(0,0)[bl]{\(15\)}}
\put(07,26){\makebox(0,0)[bl]{\(145\)}}
\put(14,26){\makebox(0,0)[bl]{\(181\)}}
\put(22.5,26){\makebox(0,0)[bl]{\(3\)}}
\put(29.5,26){\makebox(0,0)[bl]{\(5\)}}

%\put(36.5,24){\makebox(0,0)[bl]{\(1\)}}

%\put(36.5,26){\makebox(0,0)[bl]{\(2\)}}
\put(36.5,26){\makebox(0,0)[bl]{\(4\)}}
%\put(57.5,24){\makebox(0,0)[bl]{\(6\)}}

%--6
\put(01,22){\makebox(0,0)[bl]{\(16\)}}
\put(07,22){\makebox(0,0)[bl]{\(150\)}}
\put(14,22){\makebox(0,0)[bl]{\(150\)}}

%\put(07,20){\makebox(0,0)[bl]{\(170\)}}
%\put(14,20){\makebox(0,0)[bl]{\(161\)}}

\put(22.5,22){\makebox(0,0)[bl]{\(5\)}}
\put(29.5,22){\makebox(0,0)[bl]{\(7\)}}

%\put(36.5,20){\makebox(0,0)[bl]{\(1\)}}

%\put(36.5,22){\makebox(0,0)[bl]{\(2\)}}
\put(36.5,22){\makebox(0,0)[bl]{\(4\)}}
%\put(57.5,20){\makebox(0,0)[bl]{\(7\)}}

%--10->9
\put(01,18){\makebox(0,0)[bl]{\(17\)}}
\put(07,18){\makebox(0,0)[bl]{\(120\)}}
\put(14,18){\makebox(0,0)[bl]{\(140\)}}
\put(22.5,18){\makebox(0,0)[bl]{\(6\)}}
\put(29.5,18){\makebox(0,0)[bl]{\(4\)}}

%\put(36.5,16){\makebox(0,0)[bl]{\(1\)}}

%\put(36.5,18){\makebox(0,0)[bl]{\(2\)}}
\put(36.5,18){\makebox(0,0)[bl]{\(6\)}}
%\put(57.5,16){\makebox(0,0)[bl]{\(8\)}}

%--11->10
\put(01,14){\makebox(0,0)[bl]{\(18\)}}
\put(07,14){\makebox(0,0)[bl]{\(150\)}}
\put(14,14){\makebox(0,0)[bl]{\(136\)}}
\put(22.5,14){\makebox(0,0)[bl]{\(3\)}}
\put(29.5,14){\makebox(0,0)[bl]{\(6\)}}

%\put(36.5,12){\makebox(0,0)[bl]{\(1\)}}

%\put(36.5,14){\makebox(0,0)[bl]{\(2\)}}
\put(36.5,14){\makebox(0,0)[bl]{\(7\)}}
%\put(57.5,26){\makebox(0,0)[bl]{\(8\)}}

%--24->18->19
\put(01,10){\makebox(0,0)[bl]{\(19\)}}
\put(07,10){\makebox(0,0)[bl]{\(135\)}}
\put(15,10){\makebox(0,0)[bl]{\(59\)}}
\put(22.5,10){\makebox(0,0)[bl]{\(4\)}}
\put(28.5,10){\makebox(0,0)[bl]{\(13\)}}

%\put(36.5,08){\makebox(0,0)[bl]{\(1\)}}

%\put(36.5,10){\makebox(0,0)[bl]{\(3\)}}
\put(36.5,10){\makebox(0,0)[bl]{\(4\)}}
%\put(57.5,08){\makebox(0,0)[bl]{\(6\)}}

%--25->19->
\put(01,06){\makebox(0,0)[bl]{\(20\)}}
\put(07,06){\makebox(0,0)[bl]{\(147\)}}
\put(15,06){\makebox(0,0)[bl]{\(79\)}}
\put(22.5,06){\makebox(0,0)[bl]{\(5\)}}
\put(29.5,06){\makebox(0,0)[bl]{\(7\)}}

%\put(36.5,04){\makebox(0,0)[bl]{\(1\)}}

%\put(36.5,06){\makebox(0,0)[bl]{\(3\)}}
\put(35.5,06){\makebox(0,0)[bl]{\(16\)}}
%\put(57.5,04){\makebox(0,0)[bl]{\(8\)}}

%--new 25->21
\put(01,02){\makebox(0,0)[bl]{\(21\)}}
\put(07,02){\makebox(0,0)[bl]{\(127\)}}
\put(15,02){\makebox(0,0)[bl]{\(95\)}}
\put(22.5,02){\makebox(0,0)[bl]{\(5\)}}
\put(28.5,02){\makebox(0,0)[bl]{\(7\)}}

%\put(36.5,02){\makebox(0,0)[bl]{\(1\)}}

%\put(36.5,02){\makebox(0,0)[bl]{\(2\)}}
\put(36.5,02){\makebox(0,0)[bl]{\(5\)}}
%\put(57.5,02){\makebox(0,0)[bl]{\(7\)}}

%------------------------------------------------

\end{picture}
%\end{center}
%
%
%\begin{center}
\begin{picture}(55,81)
%\begin{picture}(70,80)

\put(03,00){\makebox(0,0)[bl]{Fig. 10.
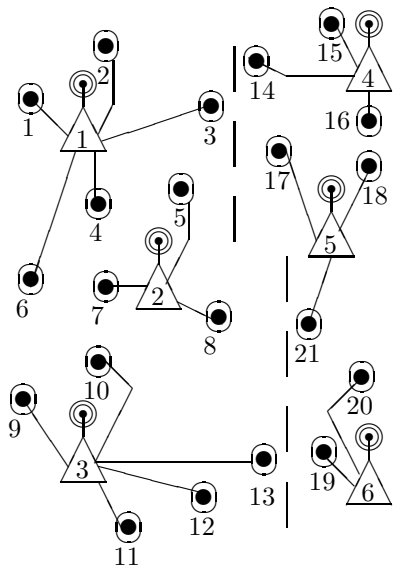
 Separated assignment \(S^{1}\)}}

%-----------------------------------------------

\put(14,32){\oval(3.2,4)} \put(14,32){\circle*{2}}
\put(12,27){\makebox(0,0)[bl]{\(10\)}}
%21->17

\put(15,42){\oval(3.2,4)} \put(15,42){\circle*{2}}
\put(13,37){\makebox(0,0)[bl]{\(7\)}}
%14->13

%------------------------------------------------Point 3 - 2
\put(19,39){\line(1,0){6}} \put(19,39){\line(1,2){3}}
\put(25,39){\line(-1,2){3}} \put(22,45){\line(0,1){3}}
\put(22,48){\circle*{1}} \put(21,39.5){\makebox(0,0)[bl]{\(2\)}}

\put(22,48){\circle{2.5}} \put(22,48){\circle{3.5}}
%--
%connection 3-15
\put(24.5,40){\line(2,-1){6}}
%--
%connection 3-8
\put(23,42.3){\line(1,2){3}} \put(26,48.3){\line(0,1){4.8}}
%--
%connection 3-13
\put(20.5,42){\line(-1,0){4.5}}

\put(05,43){\oval(3.2,4)} \put(05,43){\circle*{2}}
\put(03,38){\makebox(0,0)[bl]{\(6\)}}
%13->12

%--
%connection 1-12
\put(11,60){\line(-1,-3){6}}
%--

\put(05,67){\oval(3.2,4)} \put(05,67){\circle*{2}}
\put(03.8,62){\makebox(0,0)[bl]{\(1\)}}

\put(15,74){\oval(3.2,4)} \put(15,74){\circle*{2}}
\put(13.8,69){\makebox(0,0)[bl]{\(2\)}}

%-------------------------------------------------------Point 1
\put(09,60){\line(1,0){6}} \put(09,60){\line(1,2){3}}
\put(15,60){\line(-1,2){3}} \put(12,66){\line(0,1){3}}
\put(12,69){\circle*{1}} \put(11,60.5){\makebox(0,0)[bl]{\(1\)}}

\put(12,69){\circle{2.5}} \put(12,69){\circle{3.5}}
%--
%connection 1-1
\put(10,62){\line(-1,1){5}}
%--
%connection 1-8
\put(13.5,60){\line(0,-1){6}}
%--
%connection 1-2
\put(14,62.3){\line(1,2){2}} \put(16,66.3){\line(0,1){5.8}}

\put(14,53){\oval(3.2,4)} \put(14,53){\circle*{2}}
\put(12.8,48){\makebox(0,0)[bl]{\(4\)}}
%8->7

\put(04,27){\oval(3.2,4)} \put(04,27){\circle*{2}}
\put(02,22){\makebox(0,0)[bl]{\(9\)}}
%20->16

\put(18,10){\oval(3.2,4)} \put(18,10){\circle*{2}}
\put(16,5){\makebox(0,0)[bl]{\(11\)}}
%28->20

\put(28,14){\oval(3.2,4)} \put(28,14){\circle*{2}}
\put(26,9){\makebox(0,0)[bl]{\(12\)}}
%29->21

\put(30,38){\oval(3.2,4)} \put(30,38){\circle*{2}}
\put(28,33){\makebox(0,0)[bl]{\(8\)}}
%15->14

\put(38,60){\oval(3.2,4)} \put(38,60){\circle*{2}}
\put(36,55){\makebox(0,0)[bl]{\(17\)}}
%10->9

\put(50,58){\oval(3.2,4)} \put(50,58){\circle*{2}}
\put(49,53){\makebox(0,0)[bl]{\(18\)}}
%11->10

%-------------------------------------------------------Point 4 -5
\put(42,46){\line(1,0){6}} \put(42,46){\line(1,2){3}}
\put(48,46){\line(-1,2){3}} \put(45,52){\line(0,1){3}}
\put(45,55){\circle*{1}} \put(44,46.5){\makebox(0,0)[bl]{\(5\)}}

\put(45,55){\circle{2.5}} \put(45,55){\circle{3.5}}
%--
%connection 4-9
\put(43,48){\line(-1,3){4}}

%--
%connection 4-11
\put(46,49.4){\line(1,2){4.5}}
%--

\put(49,30){\oval(3.2,4)} \put(49,30){\circle*{2}}
\put(47,25){\makebox(0,0)[bl]{\(20\)}}
%25->19->20

\put(44,20){\oval(3.2,4)} \put(44,20){\circle*{2}}
\put(42,15){\makebox(0,0)[bl]{\(19\)}}
%24->18->19

\put(42,37){\oval(3.2,4)} \put(42,37){\circle*{2}}
\put(40,32){\makebox(0,0)[bl]{\(21\)}}
%%25->21

%--
%connection 4-25
\put(42,37){\line(1,3){3}}

%---------------------------------------------------------Point 6
\put(47,13){\line(1,0){6}} \put(47,13){\line(1,2){3}}
\put(53,13){\line(-1,2){3}} \put(50,19){\line(0,1){3}}
\put(50,22){\circle*{1}} \put(49,13.5){\makebox(0,0)[bl]{\(6\)}}

\put(50,22){\circle{2.5}} \put(50,22){\circle{3.5}}
%--
%connection 6-24
%\put(61,20){\line(-2,-1){9}}
%--
%connection 6-18
\put(48,15.5){\line(-1,1){4.5}}
%--
%connection 6-18
\put(48.7,17){\line(-1,2){4.2}} \put(44.5,25.4){\line(1,1){4}}
%--

\put(25,55){\oval(3.2,4)} \put(25,55){\circle*{2}}
\put(24,50){\makebox(0,0)[bl]{\(5\)}}
%9->8
%%%%%%%%%%%%%%%%%%%%%%%%%%%%%%%%%%%%%%%%%%%%%%%%%%%%%%%%%%%%%%%%
\put(29,66){\oval(3.2,4)} \put(29,66){\circle*{2}}
\put(28,61){\makebox(0,0)[bl]{\(3\)}}
%--
%connection 1-3
\put(28.5,66){\line(-3,-1){14}}

\put(35,72){\oval(3.2,4)} \put(35,72){\circle*{2}}
\put(34,67){\makebox(0,0)[bl]{\(14\)}}

\put(45,77){\oval(3.2,4)} \put(45,77){\circle*{2}}
\put(43,72){\makebox(0,0)[bl]{\(15\)}}

%%>>>>>>>>>>>>>>>>>>>>>>>>>>>>>>>>>>>>>>>>>>>>>>>>>>>>>>>>>>>>>>>>>>>>>>>>>>>>>>>>>>>>
\put(50,64){\oval(3.2,4)} \put(50,64){\circle*{2}}
\put(44,63){\makebox(0,0)[bl]{\(16\)}}

%connection 4-16
\put(50,64){\line(0,1){4}}

%---------------------------------------------------Point 2 -4
\put(47,68){\line(1,0){6}} \put(47,68){\line(1,2){3}}
\put(53,68){\line(-1,2){3}} \put(50,74){\line(0,1){3}}
\put(50,77){\circle*{1}} \put(49,68.5){\makebox(0,0)[bl]{\(4\)}}

\put(50,77){\circle{2.5}} \put(50,77){\circle{3.5}}
%--
%connection 2-4
\put(48,70){\line(-1,0){09}} \put(39,70){\line(-2,1){5}}
%--
%connection 2-5
\put(48.4,71.2){\line(-1,2){3}}
%--

%---------------------------------------------------------Point 5-3
\put(09,16){\line(1,0){6}} \put(09,16){\line(1,2){3}}
\put(15,16){\line(-1,2){3}} \put(12,22){\line(0,1){3}}
\put(12,25){\circle*{1}} \put(11,16.5){\makebox(0,0)[bl]{\(3\)}}

\put(12,25){\circle{2.5}} \put(12,25){\circle{3.5}}
%------------------------------------------------
%--
%connection 5-20
\put(10,18){\line(-2,3){6.5}}
%--
%connection 5-21
\put(13.5,18.5){\line(1,2){05}} \put(18.5,28.5){\line(-1,1){04}}
%--
%connection 5-29
\put(14,18){\line(4,-1){14}}
%--
%connection 5-28
\put(14,16){\line(1,-2){3}}
%--

%%%%%%%%%%%%%%%%%%%%%%%%%%%%%%%%%%%%%%%%%%%%%%%%%%%%%%%%%%%%%%%%%%
%%%%%%%%%%%%%%%%%%%%%%%%%%%%%%%%%%%%%%%%%%%%%%%%%%%%%%%% Additions

\put(36,19){\oval(3.2,4)} \put(36,19){\circle*{2}}
\put(34,13){\makebox(0,0)[bl]{\(13\)}}
%connection 5-23
\put(34.4,18.6){\line(-1,0){20.6}}
%--

%%%%%%%%%%%%%%%%%%%%%%%%%%%%%%%%%%%%% border
\put(39,10){\line(0,1){6}} \put(39.1,10){\line(0,1){6}}
%-
\put(39,20){\line(0,1){6}} \put(39.1,20){\line(0,1){6}}
%-
\put(39,30){\line(0,1){6}} \put(39.1,30){\line(0,1){6}}
%-
\put(39,40){\line(0,1){6}} \put(39.1,40){\line(0,1){6}}
%-
\put(32,48){\line(0,1){6}} \put(32.1,48){\line(0,1){6}}
%-
\put(32,58){\line(0,1){6}} \put(32.1,58){\line(0,1){6}}
%-
\put(32,68){\line(0,1){6}} \put(32.1,68){\line(0,1){6}}
%-
%%%%%%%%%%%%%%%%%%%%%%%%%%%%%%%%%%%%%%%%%%%%%%%%%%%%%%%%%%

\end{picture}
\end{center}

\begin{center}
%\begin{picture}(55,80)
\begin{picture}(60,81)

\put(05,00){\makebox(0,0)[bl]{Fig. 11.
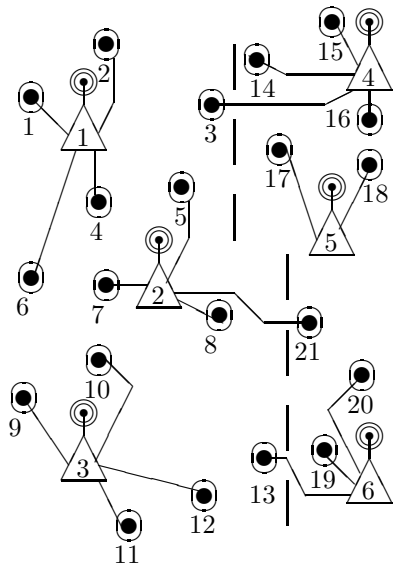
 Joint assignment \(S^{2}\)}}

%-----------------------------------------------

\put(14,32){\oval(3.2,4)} \put(14,32){\circle*{2}}
\put(12,27){\makebox(0,0)[bl]{\(10\)}}
%21->17

\put(15,42){\oval(3.2,4)} \put(15,42){\circle*{2}}
\put(13,37){\makebox(0,0)[bl]{\(7\)}}
%14->13

%------------------------------------------------Point 3-2
\put(19,39){\line(1,0){6}} \put(19,39){\line(1,2){3}}
\put(25,39){\line(-1,2){3}} \put(22,45){\line(0,1){3}}
\put(22,48){\circle*{1}} \put(21,39.5){\makebox(0,0)[bl]{\(2\)}}

\put(22,48){\circle{2.5}} \put(22,48){\circle{3.5}}
%---
%Connection 3-10, 3-15, 3-25, 3-13
%--
%connection 3-15
\put(24.5,40){\line(2,-1){6}}
%--
%connection 3-8
\put(23,42.3){\line(1,2){3}} \put(26,48.3){\line(0,1){4.8}}
%--
%connection 3-13
\put(20.5,42){\line(-1,0){4.5}}

\put(05,43){\oval(3.2,4)} \put(05,43){\circle*{2}}
\put(03,38){\makebox(0,0)[bl]{\(6\)}}
%13->12

%--
%connection 1-12
\put(11,60){\line(-1,-3){6}}
%--

\put(05,67){\oval(3.2,4)} \put(05,67){\circle*{2}}
\put(03.8,62){\makebox(0,0)[bl]{\(1\)}}

\put(15,74){\oval(3.2,4)} \put(15,74){\circle*{2}}
\put(13.8,69){\makebox(0,0)[bl]{\(2\)}}

%-------------------------------------------------------Point 1
\put(09,60){\line(1,0){6}} \put(09,60){\line(1,2){3}}
\put(15,60){\line(-1,2){3}} \put(12,66){\line(0,1){3}}
\put(12,69){\circle*{1}} \put(11,60.5){\makebox(0,0)[bl]{\(1\)}}

\put(12,69){\circle{2.5}} \put(12,69){\circle{3.5}}
%---
%Connection 1-1, 1-2, 1-8, 1-12
%--
%connection 1-1
\put(10,62){\line(-1,1){5}}
%--
%connection 1-8
\put(13.5,60){\line(0,-1){6}}
%--
%connection 1-2
\put(14,62.3){\line(1,2){2}} \put(16,66.3){\line(0,1){5.8}}

\put(14,53){\oval(3.2,4)} \put(14,53){\circle*{2}}
\put(12.8,48){\makebox(0,0)[bl]{\(4\)}}
%8->7

\put(04,27){\oval(3.2,4)} \put(04,27){\circle*{2}}
\put(02,22){\makebox(0,0)[bl]{\(9\)}}
%20->16

\put(18,10){\oval(3.2,4)} \put(18,10){\circle*{2}}
\put(16,5){\makebox(0,0)[bl]{\(11\)}}
%28->20

\put(28,14){\oval(3.2,4)} \put(28,14){\circle*{2}}
\put(26,9){\makebox(0,0)[bl]{\(12\)}}
%29->21

\put(30,38){\oval(3.2,4)} \put(30,38){\circle*{2}}
\put(28,33){\makebox(0,0)[bl]{\(8\)}}
%15->14

\put(38,60){\oval(3.2,4)} \put(38,60){\circle*{2}}
\put(36,55){\makebox(0,0)[bl]{\(17\)}}
%10->9

\put(50,58){\oval(3.2,4)} \put(50,58){\circle*{2}}
\put(49,53){\makebox(0,0)[bl]{\(18\)}}
%11->10

%----------------------------------------------------------Point 4-5
\put(42,46){\line(1,0){6}} \put(42,46){\line(1,2){3}}
\put(48,46){\line(-1,2){3}} \put(45,52){\line(0,1){3}}
\put(45,55){\circle*{1}} \put(44,46.5){\makebox(0,0)[bl]{\(5\)}}

\put(45,55){\circle{2.5}} \put(45,55){\circle{3.5}}
%---
%Connection 4-9, 4-11, 4-12
%--
%connection 4-9
%\put(43.5,48.5){\line(-3,1){4.5}} \put(39,50){\line(0,1){8}}

\put(43,48){\line(-1,3){4}}

%--
%connection 4-11
\put(46,49.4){\line(1,2){4.5}}
%--
%connection 4-12
%\put(47,48){\line(3,1){13}}
%--

%\put(60,52){\oval(3.2,4)} \put(60,52){\circle*{2}}
%\put(58,47){\makebox(0,0)[bl]{\(19\)}}
%12->11

%\put(63,35){\oval(3.2,4)} \put(63,35){\circle*{2}}
%\put(61,30){\makebox(0,0)[bl]{\(20\)}}
%19->15

\put(49,30){\oval(3.2,4)} \put(49,30){\circle*{2}}
\put(47,25){\makebox(0,0)[bl]{\(20\)}}
%25->19->20

\put(44,20){\oval(3.2,4)} \put(44,20){\circle*{2}}
\put(42,15){\makebox(0,0)[bl]{\(19\)}}
%24->18->19

\put(42,37){\oval(3.2,4)} \put(42,37){\circle*{2}}
\put(40,32){\makebox(0,0)[bl]{\(21\)}}
%%25->21

%--
%connection 4-25 => 3-25
%\put(42,37){\line(1,3){3}}

\put(42,37){\line(-1,0){6}} \put(36,37){\line(-1,1){4}}
\put(32,41){\line(-1,0){8}}

%---------------------------------------------------------Point 6
\put(47,13){\line(1,0){6}} \put(47,13){\line(1,2){3}}
\put(53,13){\line(-1,2){3}} \put(50,19){\line(0,1){3}}
\put(50,22){\circle*{1}} \put(49,13.5){\makebox(0,0)[bl]{\(6\)}}

\put(50,22){\circle{2.5}} \put(50,22){\circle{3.5}}
%---
%Connection 6-18, 6-15, 6-22, 6-19, 6-24
%--
%connection 6-24
%\put(61,20){\line(-2,-1){9}}
%--
%connection 6-18
\put(48,15.5){\line(-1,1){4.5}}
%--
%connection 6-18
\put(48.7,17){\line(-1,2){4.2}} \put(44.5,25.4){\line(1,1){4}}
%--
%connection 6-15
%\put(51,16){\line(1,2){07.5}} \put(58.5,31){\line(1,1){04}}
%--
%connection 6-22
%\put(52,15){\line(2,-1){8.6}}
%--

\put(25,55){\oval(3.2,4)} \put(25,55){\circle*{2}}
\put(24,50){\makebox(0,0)[bl]{\(5\)}}
%9->8
%%%%%%%%%%%%%%%%%%%%%%%%%%%%%%%%%%%%%%%%%%%%%%%%%%%%%%%%%%%%%%%%
\put(29,66){\oval(3.2,4)} \put(29,66){\circle*{2}}
\put(28,61){\makebox(0,0)[bl]{\(3\)}}
%--
%connection 1-3 => 2-3
\put(29,66){\line(1,0){15}} \put(44,66){\line(2,1){4}}

\put(35,72){\oval(3.2,4)} \put(35,72){\circle*{2}}
\put(34,67){\makebox(0,0)[bl]{\(14\)}}

\put(45,77){\oval(3.2,4)} \put(45,77){\circle*{2}}
\put(43,72){\makebox(0,0)[bl]{\(15\)}}

%%>>>>>>>>>>>>>>>>>>>>>>>>>>>>>>>>>>>>>>>>>>>>>>>>>>>>>>>>>>>>>>>>>>>>>>>>>>>>>>>>>>>>
\put(50,64){\oval(3.2,4)} \put(50,64){\circle*{2}}
\put(44,63){\makebox(0,0)[bl]{\(16\)}}

%connection 4-16
\put(50,64){\line(0,1){4}}

%----------------------------------------------------------Point 2-4
\put(47,68){\line(1,0){6}} \put(47,68){\line(1,2){3}}
\put(53,68){\line(-1,2){3}} \put(50,74){\line(0,1){3}}
\put(50,77){\circle*{1}} \put(49,68.5){\makebox(0,0)[bl]{\(4\)}}

\put(50,77){\circle{2.5}} \put(50,77){\circle{3.5}}
%--
%connection 2-4
\put(48,70){\line(-1,0){09}} \put(39,70){\line(-2,1){5}}
%--
%connection 2-5
\put(48.4,71.2){\line(-1,2){3}}
%--
%connection 2-6
%%%%%%\put(52,70){\line(2,-1){5}}
%--

%---------------------------------------------------------Point 5-3
\put(09,16){\line(1,0){6}} \put(09,16){\line(1,2){3}}
\put(15,16){\line(-1,2){3}} \put(12,22){\line(0,1){3}}
\put(12,25){\circle*{1}} \put(11,16.5){\makebox(0,0)[bl]{\(3\)}}

\put(12,25){\circle{2.5}} \put(12,25){\circle{3.5}}
%------------------------------------------------
%Connection 5-20, 5-21, 5-29, 5-28
%--
%connection 5-20
\put(10,18){\line(-2,3){6.5}}
%--
%connection 5-21
\put(13.5,18.5){\line(1,2){05}} \put(18.5,28.5){\line(-1,1){04}}
%--
%connection 5-29
\put(14,18){\line(4,-1){14}}
%--
%connection 5-28
\put(14,16){\line(1,-2){3}}
%--

%%%%%%%%%%%%%%%%%%%%%%%%%%%%%%%%%%%%%%%%%%%%%%%%%%%%%%%%%%%%%%%%%%
%%%%%%%%%%%%%%%%%%%%%%%%%%%%%%%%%%%%%%%%%%%%%%%%%%%%%%%% Additions

\put(36,19){\oval(3.2,4)} \put(36,19){\circle*{2}}
\put(34,13){\makebox(0,0)[bl]{\(13\)}}
%connection 5-23 6-23
%\put(34.4,18.6){\line(-1,0){20.6}}

\put(36,19){\line(1,0){3}} \put(39,19){\line(1,-2){2.5}}
\put(41.5,14){\line(1,0){6}}

%--

%%%%%%%%%%%%%%%%%%%%%%%%%%%%%%%%%%%%% border
\put(39,10){\line(0,1){6}} \put(39.1,10){\line(0,1){6}}
%-
\put(39,20){\line(0,1){6}} \put(39.1,20){\line(0,1){6}}
%-
\put(39,30){\line(0,1){6}} \put(39.1,30){\line(0,1){6}}
%-
\put(39,40){\line(0,1){6}} \put(39.1,40){\line(0,1){6}}
%-
\put(32,48){\line(0,1){6}} \put(32.1,48){\line(0,1){6}}
%-
\put(32,58){\line(0,1){6}} \put(32.1,58){\line(0,1){6}}
%-
\put(32,68){\line(0,1){6}} \put(32.1,68){\line(0,1){6}}
%-
%%%%%%%%%%%%%%%%%%%%%%%%%%%%%%%%%%%%%%%%%%%%%%%%%%%%%%%%%%

\end{picture}
%\end{center}
%
%\begin{center}
\begin{picture}(55,81)
\put(05,00){\makebox(0,0)[bl]{Fig. 12.
 Joint assignment \(S^{*}\)}}

%-----------------------------------------------

\put(14,32){\oval(3.2,4)} \put(14,32){\circle*{2}}
\put(12,27){\makebox(0,0)[bl]{\(10\)}}
%21->17

\put(15,42){\oval(3.2,4)} \put(15,42){\circle*{2}}
\put(13,37){\makebox(0,0)[bl]{\(7\)}}
%14->13

%------------------------------------------------Point 3-2
\put(19,39){\line(1,0){6}} \put(19,39){\line(1,2){3}}
\put(25,39){\line(-1,2){3}} \put(22,45){\line(0,1){3}}
\put(22,48){\circle*{1}} \put(21,39.5){\makebox(0,0)[bl]{\(2\)}}

\put(22,48){\circle{2.5}} \put(22,48){\circle{3.5}}
%---
%Connection 3-10, 3-15, 3-25, 3-13
%--
%connection 3-15
\put(24.5,40){\line(2,-1){6}}
%--
%connection 3-8
\put(23,42.3){\line(1,2){3}} \put(26,48.3){\line(0,1){4.8}}
%--
%connection 3-13
\put(20.5,42){\line(-1,0){4.5}}

\put(05,43){\oval(3.2,4)} \put(05,43){\circle*{2}}
\put(03,38){\makebox(0,0)[bl]{\(6\)}}
%13->12

%--
%connection 1-12
\put(11,60){\line(-1,-3){6}}
%--

\put(05,67){\oval(3.2,4)} \put(05,67){\circle*{2}}
\put(03.8,62){\makebox(0,0)[bl]{\(1\)}}

\put(15,74){\oval(3.2,4)} \put(15,74){\circle*{2}}
\put(13.8,69){\makebox(0,0)[bl]{\(2\)}}

%-------------------------------------------------------Point 1
\put(09,60){\line(1,0){6}} \put(09,60){\line(1,2){3}}
\put(15,60){\line(-1,2){3}} \put(12,66){\line(0,1){3}}
\put(12,69){\circle*{1}} \put(11,60.5){\makebox(0,0)[bl]{\(1\)}}

\put(12,69){\circle{2.5}} \put(12,69){\circle{3.5}}
%---
%Connection 1-1, 1-2, 1-8, 1-12
%--
%connection 1-1
\put(10,62){\line(-1,1){5}}
%--
%connection 1-8
\put(13.5,60){\line(0,-1){6}}
%--
%connection 1-2
\put(14,62.3){\line(1,2){2}} \put(16,66.3){\line(0,1){5.8}}

\put(14,53){\oval(3.2,4)} \put(14,53){\circle*{2}}
\put(12.8,48){\makebox(0,0)[bl]{\(4\)}}
%8->7

\put(04,27){\oval(3.2,4)} \put(04,27){\circle*{2}}
\put(02,22){\makebox(0,0)[bl]{\(9\)}}
%20->16

\put(18,10){\oval(3.2,4)} \put(18,10){\circle*{2}}
\put(16,5){\makebox(0,0)[bl]{\(11\)}}
%28->20

\put(28,14){\oval(3.2,4)} \put(28,14){\circle*{2}}
\put(26,9){\makebox(0,0)[bl]{\(12\)}}
%29->21

\put(30,38){\oval(3.2,4)} \put(30,38){\circle*{2}}
\put(28,33){\makebox(0,0)[bl]{\(8\)}}
%15->14

\put(38,60){\oval(3.2,4)} \put(38,60){\circle*{2}}
\put(36,55){\makebox(0,0)[bl]{\(17\)}}
%10->9

\put(50,58){\oval(3.2,4)} \put(50,58){\circle*{2}}
\put(49,53){\makebox(0,0)[bl]{\(18\)}}
%11->10

%%%%%%%%%%%%%%%%%%%%%%%% New-Nov5,2010
%connection 1-3
\put(28.5,66){\line(-3,-1){14}}

%----------------------------------------------------------Point 4-5
\put(42,46){\line(1,0){6}} \put(42,46){\line(1,2){3}}
\put(48,46){\line(-1,2){3}} \put(45,52){\line(0,1){3}}
\put(45,55){\circle*{1}} \put(44,46.5){\makebox(0,0)[bl]{\(5\)}}

\put(45,55){\circle{2.5}} \put(45,55){\circle{3.5}}
%---
%Connection 4-9, 4-11, 4-12
%--
%connection 4-9
%\put(43.5,48.5){\line(-3,1){4.5}} \put(39,50){\line(0,1){8}}

\put(43,48){\line(-1,3){4}}

%--
%connection 4-11
\put(46,49.4){\line(1,2){4.5}}
%--
%connection 4-12
%\put(47,48){\line(3,1){13}}
%--

%\put(60,52){\oval(3.2,4)} \put(60,52){\circle*{2}}
%\put(58,47){\makebox(0,0)[bl]{\(19\)}}
%12->11

%\put(63,35){\oval(3.2,4)} \put(63,35){\circle*{2}}
%\put(61,30){\makebox(0,0)[bl]{\(20\)}}
%19->15

\put(49,30){\oval(3.2,4)} \put(49,30){\circle*{2}}
\put(47,25){\makebox(0,0)[bl]{\(20\)}}
%25->19->22->20

\put(44,20){\oval(3.2,4)} \put(44,20){\circle*{2}}
\put(42,15){\makebox(0,0)[bl]{\(19\)}}
%24->18->21->19

\put(42,37){\oval(3.2,4)} \put(42,37){\circle*{2}}
\put(40,32){\makebox(0,0)[bl]{\(21\)}}
%%25->21

%--
%connection 4-25 => 3-25
%\put(42,37){\line(1,3){3}}

\put(42,37){\line(-1,0){6}} \put(36,37){\line(-1,1){4}}
\put(32,41){\line(-1,0){8}}

%---------------------------------------------------------Point 6
\put(47,13){\line(1,0){6}} \put(47,13){\line(1,2){3}}
\put(53,13){\line(-1,2){3}} \put(50,19){\line(0,1){3}}
\put(50,22){\circle*{1}} \put(49,13.5){\makebox(0,0)[bl]{\(6\)}}

\put(50,22){\circle{2.5}} \put(50,22){\circle{3.5}}
%---
%Connection 6-18, 6-15, 6-22, 6-19, 6-24
%--
%connection 6-24
%\put(61,20){\line(-2,-1){9}}
%--
%connection 6-18
\put(48,15.5){\line(-1,1){4.5}}
%--
%connection 6-18
\put(48.7,17){\line(-1,2){4.2}} \put(44.5,25.4){\line(1,1){4}}
%--
%connection 6-15
%\put(51,16){\line(1,2){07.5}} \put(58.5,31){\line(1,1){04}}
%--
%connection 6-22
%\put(52,15){\line(2,-1){8.6}}
%--

\put(25,55){\oval(3.2,4)} \put(25,55){\circle*{2}}
\put(24,50){\makebox(0,0)[bl]{\(5\)}}
%9->8
%%%%%%%%%%%%%%%%%%%%%%%%%%%%%%%%%%%%%%%%%%%%%%%%%%%%%%%%%%%%%%%%
\put(29,66){\oval(3.2,4)} \put(29,66){\circle*{2}}
\put(28,61){\makebox(0,0)[bl]{\(3\)}}
%--
%connection 1-3 => 2-3
%%%%%%%%%%%%%%%%%%%\put(29,66){\line(1,0){15}} \put(44,66){\line(2,1){4}}

\put(35,72){\oval(3.2,4)} \put(35,72){\circle*{2}}
\put(34,67){\makebox(0,0)[bl]{\(14\)}}

\put(45,77){\oval(3.2,4)} \put(45,77){\circle*{2}}
\put(43,72){\makebox(0,0)[bl]{\(15\)}}

%%>>>>>>>>>>>>>>>>>>>>>>>>>>>>>>>>>>>>>>>>>>>>>>>>>>>>>>>>>>>>>>>>>>>>>>>>>>>>>>>>>>>>
\put(50,64){\oval(3.2,4)} \put(50,64){\circle*{2}}
\put(44,63){\makebox(0,0)[bl]{\(16\)}}

%connection 4-16
\put(50,64){\line(0,1){4}}

%----------------------------------------------------------Point 2-4
\put(47,68){\line(1,0){6}} \put(47,68){\line(1,2){3}}
\put(53,68){\line(-1,2){3}} \put(50,74){\line(0,1){3}}
\put(50,77){\circle*{1}} \put(49,68.5){\makebox(0,0)[bl]{\(4\)}}

\put(50,77){\circle{2.5}} \put(50,77){\circle{3.5}}
%--
%connection 2-4
\put(48,70){\line(-1,0){09}} \put(39,70){\line(-2,1){5}}
%--
%connection 2-5
\put(48.4,71.2){\line(-1,2){3}}
%--

%---------------------------------------------------------Point 5-3
\put(09,16){\line(1,0){6}} \put(09,16){\line(1,2){3}}
\put(15,16){\line(-1,2){3}} \put(12,22){\line(0,1){3}}
\put(12,25){\circle*{1}} \put(11,16.5){\makebox(0,0)[bl]{\(3\)}}

\put(12,25){\circle{2.5}} \put(12,25){\circle{3.5}}
%------------------------------------------------
%Connection 5-20, 5-21, 5-29, 5-28
%--
%connection 5-20
\put(10,18){\line(-2,3){6.5}}
%--
%connection 5-21
\put(13.5,18.5){\line(1,2){05}} \put(18.5,28.5){\line(-1,1){04}}
%--
%connection 5-29
\put(14,18){\line(4,-1){14}}
%--
%connection 5-28
\put(14,16){\line(1,-2){3}}
%--

%%%%%%%%%%%%%%%%%%%%%%%%%%%%%%%%%%%%%%%%%%%%%%%%%%%%%%%%%%%%%%%%%%
%%%%%%%%%%%%%%%%%%%%%%%%%%%%%%%%%%%%%%%%%%%%%%%%%%%%%%%% Additions

\put(36,19){\oval(3.2,4)} \put(36,19){\circle*{2}}
\put(34,13){\makebox(0,0)[bl]{\(13\)}}
%connection 5-23 6-23
%\put(34.4,18.6){\line(-1,0){20.6}}

\put(36,19){\line(1,0){3}} \put(39,19){\line(1,-2){2.5}}
\put(41.5,14){\line(1,0){6}}

%--

%%%%%%%%%%%%%%%%%%%%%%%%%%%%%%%%%%%%% border
\put(39,10){\line(0,1){6}} \put(39.1,10){\line(0,1){6}}
%-
\put(39,20){\line(0,1){6}} \put(39.1,20){\line(0,1){6}}
%-
\put(39,30){\line(0,1){6}} \put(39.1,30){\line(0,1){6}}
%-
\put(39,40){\line(0,1){6}} \put(39.1,40){\line(0,1){6}}
%-
\put(32,48){\line(0,1){6}} \put(32.1,48){\line(0,1){6}}
%-
\put(32,58){\line(0,1){6}} \put(32.1,58){\line(0,1){6}}
%-
\put(32,68){\line(0,1){6}} \put(32.1,68){\line(0,1){6}}
%-
%%%%%%%%%%%%%%%%%%%%%%%%%%%%%%%%%%%%%%%%%%%%%%%%%%%%%%%%%%

\end{picture}
\end{center}

\section{Conclusion}

 In the paper
 a restructuring approach in combinatorial optimization is suggested.
 The restructuring problem is formulated as a combinatorial
 optimization problem with one objective function.
 Multicriteria problem statement is briefly described as well.
 The restructuring approach is applied
 for several combinatorial optimization problems
 (knapsack problem, multiple choice problem,
 assignment problem, minimum spanning tree, Steiner tree problem).
 Some application domains are pointed out
 (e.g., sensors, communication networks).
% , distributed computer systems).
%
%%%%%%%%%%%%%%%%%%%%%%%%%%%%%%%%%%%%%%%%%%%%%%%%%%%%%%%%%%%%%%%%%%%%%%%%%%%%%%
%
 The suggested restructuring approach
 is the first step in this research field.
 Clearly, it is reasonable to consider other types of
  system reconfiguration problems.
% (e.g., \cite{lev09}, \cite{vrba10}).
%
%
 In the future it may be prospective
  to consider the following research directions:
 {\it 1.} application of the suggested restructuring approach to
 other combinatorial optimization problems
 (e.g., covering,
% scheduling,
 graph coloring);
 {\it 2.} examination of multicriteria restructuring models;
 {\it 3.} examination of restructuring problems with changes of basic element sets
 (i.e., \(A^{1} \neq  A^{2}\));
 {\it 4.} study and usage of various types of proximity between
 obtained solution(s) and goal solution(s)
 (i.e., \( \rho (S^{*},S^{2})\));
% and corresponding restructuring problems;
%
% of designing a system improvement trajectory
%   (i.e., multistage system redesign);
%
% {\it 1.} examination of designing a system improvement trajectory
%   (i.e., multistage system redesign);
%
%  {\it 2.} analysis of on-line redesign/adaption
%  problems for building automation  systems;
%
  {\it 5.} examination of the restructuring problems under uncertainty
  (e.g., stochastic models, fuzzy sets based models);
%
% ~{\it 5.} analysis of dynamical problems
% (e.g., as multistage design);
%
 {\it 6.} reformulation of restructuring problem(s)
 as satisfiability model(s);
 {\it 7.} usage of various AI techniques in solving procedures;
 and
 {\it 8.} application of the suggested restructuring approaches in engineering/CS education.

%%%%%%%%%%%%%%%%%%%%%%%%%%%%%%%%%%%%%%%%%%%%%%%%%%%%%%%%%%%%%%%%%%%%%%%%%%

\end{document}